\documentclass[12pt]{article}

\usepackage{graphicx}

\suppressfloats[!]

\begin{document}
        
\title{  Monte-Carlo simulation of lepton pair production in
$\bar{p}p \rightarrow l^{+}l^{-} + X$ events at 
$E_{beam}$ = 14 GeV}

\author{  A.N.~Skachkova, N.B.~Skachkov, G.D. Alexeev,\\
Joint Institute for Nuclear Research,
Dubna, Russia\\
E-mail:   Anna.Skachkova@cern.ch, skachkov@jinr.ru, alexeev@jinr.ru}

\maketitle


\begin{abstract}
\noindent 
~~     The lepton pair production 
    in collisions  of antiproton beam ($E_{beam}$ = 14 GeV)
    with proton target is studied  on the basis of event 
    samples  simulated with PYTHIA6 generator. Two types of 
    quark level subprocesses are considered. The first one 
    goes through the  production of virtual photon which
    converts into lepton pair 
    ${q\bar{q} \to \gamma^{*} \to l^{+}l^{-}}$ 
    having a continuous energy spectrum of the final lepton pair
    invariant mass. The other subprocess  proceeds through the
     $J/\Psi $ resonance production 
    $p + \bar p \to J/\Psi  + X \to l^{+}l^{-} + X$
    with the following decay of  J/$\Psi$ into a pair of 
    leptons. The distributions of  different kinematical variables
    which may be useful for the  design  of the  muon system and the 
    electromagnetic calorimeter  of  the detector of 
    PANDA experiment at FAIR  are presented.
     The analysis of these distribution shows the
    possibility to measure  the proton structure function
    in a new kinematical region defined by the time-like values of the
    square of the momentum transferred
    $1 < Q^{2} <6.25$ GeV$^{2}$ and withing a rather wide
    interval  $0.05 < x < 0.7$  of Bjorken x-variable. 
    It is also  argued that the measurement of the total transverse
    momentum  of a lepton-antilepton system may provide 
    important information about the intrinsic transverse 
    momentum $<k_{T}>$ that appears due to the 
    Fermi motion  of quarks inside the nucleon. Another 
    interesting possibility 
    is the measurement of the production rate
    of two or three lepton pairs in one 
    $ \bar pp$-collision  event that can 
    give the information about the rate of multiple quark
    interactions and the proton space structure.
        The problems due to the presence of fake  leptons 
    that  appear from meson decays, as well as due to   the
    background caused by minimum bias events and
    other QCD processes, are also discussed. The set of cuts 
    which allows one to separate the signal events  with
    lepton pairs from this kind of background events is proposed.

\end{abstract}

\newpage
\tableofcontents
\newpage
\section{Introduction}
\label{intro}

~~~~ The measurements of lepton pair production
   in hadron-hadron interactions (in the following, MMTDY
   process, see \cite{Matv}  and \cite{Drell}) have already demonstrated their great potential for studying the
   properties of elementary particles. As for
   illustration, it is enough to mention the facts 
   of  discovery of charmed  J(J/$\Psi$) - meson 
   as well as of beauty $\Upsilon$ - meson which
   were done first in hadron-hadron collisions 
   and  confirmed later  in  $e^{+}e^{-}$ experiments. 
   Dilepton events may serve  as a powerful 
   tool to get out the information about the 
   parton distribution  functions (PDFs) in
   hadrons as it was already  shown in 
   a number of high energy experiments \cite{DYexper}
   and  theoretical papers, devoted to the data
   analysis in the  framework of QCD \cite{MRST}. 
   The plans to study  this process are included 
   into the LoI \cite{PANDALoI}, TPR  \cite{PANDATPR} and \cite{PANPysBook} of PANDA experiment at HESR 
   \footnote {Analogous arguments in a favor of 
   studying of this process may be found also in 
   a number of recent proposals for experimental
   program at HESR (see \cite{ASSIA}, \cite{PAX}).} 
   which  may  provide an  interesting information 
   about quark dynamics inside the nucleon.

   This intermediate energy experiment  (in the
   following we shall consider the case of 
   antiproton beam energy $E_{beam}=14$ GeV which
   corresponds  to the center-of-mass energy
   of the $p\bar{p}$ system  $E_{cm}= 5.3$  GeV)
   may play an important role because it allows one
   to study the energy range where the perturbative 
   methods of QCD (pQCD) come into interplay
   with a rich physics of bound states and 
   resonances.  The physics of hadron resonances
   formation and decay  is strongly connected with 
   the confinement problem, i.e. with the parton
   dynamics at large distances.
    A  detailed and  high-precision  experimental 
   study at PANDA may allow one to discriminate 
   between  a large variety of existing  nonperturbative 
   approaches and models that already exist or 
   are under  development now.

     To reach the goals declared in the \cite{PANDALoI}, 
     \cite{PANDATPR}, \cite{PANPysBook} in  
     connection with lepton pair production process 
     \cite{Matv}, \cite{Drell}, one needs to know  the
     possible energy, momentum and angle distributions
     of   the produced individual leptons as well as the
     analogous distributions for the lepton pair as a 
     whole. So, a detailed Monte-Carlo simulation of 
     $\bar{p}p \rightarrow l^{+}l^{-} + X$ 
     {\bf($l = \mu, e$)} process 
     (see Fig.1) is needed. It is also clear that 
     such a sort of simulation is also needed for 
     a proper design of the muon system and 
     electromagnetic calorimeter (see \cite{Note1},
     \cite{Note2}).

   \begin{figure}[!ht]
     \begin{center}
    \mbox{\includegraphics[width=12cm, height=7.2cm]{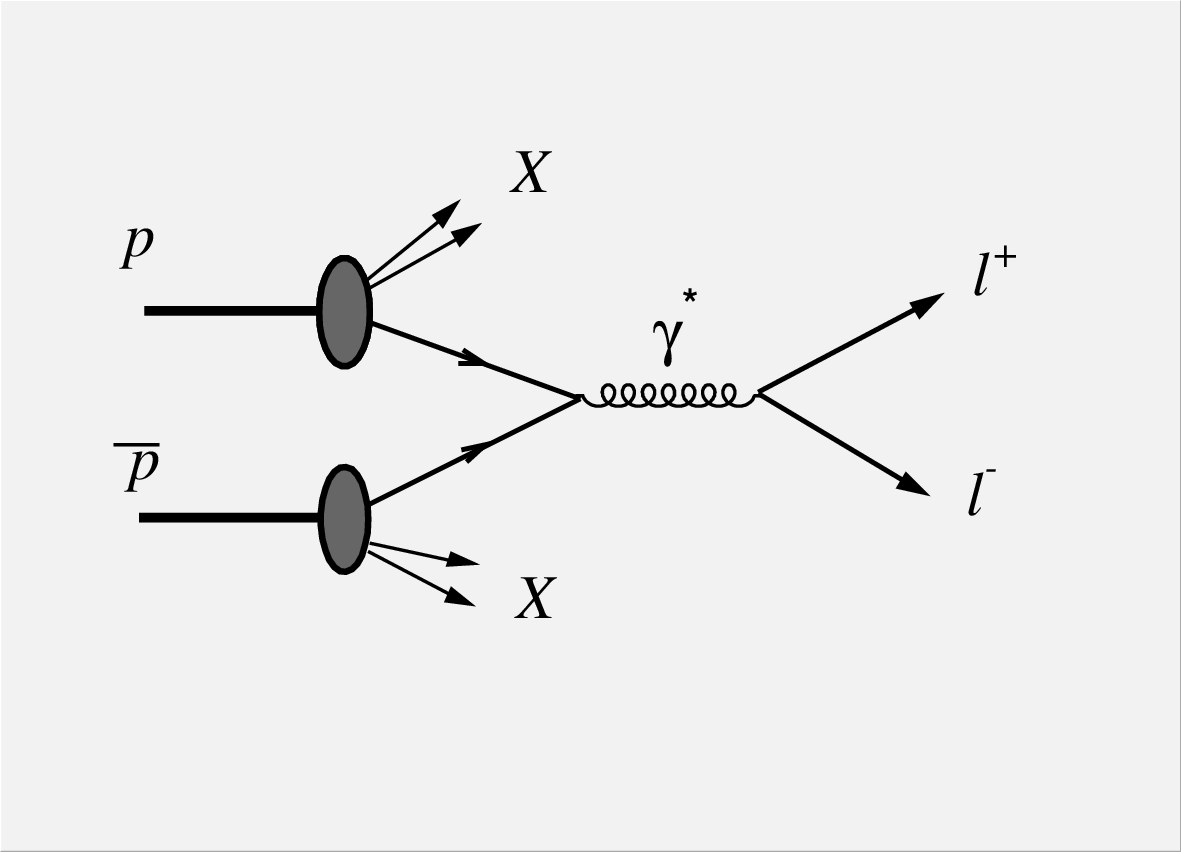}}
     \caption{\small \it $\bar{p}p \rightarrow 
     l^{+}l^{-} + X$ process }
     \end{center}
     \end{figure}

    For this aim we utilized here, as for the first
   step, the well known  event generator  
   PYTHIA  \cite{Sjost},  which is based on the ideas of 
   the quark parton model  and is well tested  and 
   widely used for the simulation of  hadron-hadron
   interactions. PYTHIA simulation is based on the 
   use of the amplitudes of the  relativistic 
   quantum field theory. This allows a proper account of the
   relativistic kinematics during simulation of different physical
   variables  distributions specific for 
   $\mu^{+}\mu^{-}$- or  $e^{+}e^{-}$- pair 
   produced in event. In our case  we use the perturbative QCD/QED 
   parton level amplitude  of the lepton pair production  process
    ${q\bar{q} \to \gamma^{*} \to l^{+}l^{-}}$   with the continuous
   spectrum of the invariant mass of lepton pair
   and the amplitude of J/$\Psi$ - resonance production process
   $p + \bar p \to J/\Psi  + X \to l^{+}l^{-} + X$,
   where J/$\Psi$ decays through the leptonic channel
   ($J/\Psi \to l^{+}l^{-}$, $l = \mu, e$ ),
   which are  implemented into the PYTHIA package.
 
   Let us underline that the results obtained here
   on the basis of PYTHIA simulation in some sense
   may allow to fix the predictions for lepton
   kinematical distributions which  may be obtained 
   in the framework of perturbative theory approach. 
   Therefore, they may be useful at the analysis 
   stage for defining the boundary between  the
   predictions of   perturbative and 
   nonperturbative  theoretical approaches.

    In Section 2 we present the kinematical 
   distributions for individual leptons. The set 
   of plots with energy, transverse momentum and 
   angle distributions are given together  with some
    plots which show different kinds of  Energy-Energy, Energy-Angle
   and Angle-Angle correlations between the physical
   variables of the leptons produced via the leading order  quark
   level subprocess $q\bar{q} \rightarrow l^{+}l^{-}$.
   The estimations of events loss due to {\bf a}  possible
   different choice of geometrical parameters of
   muon system and electromagnetic calorimeter are
   given.

    Section 3 is devoted to another channel of 
    lepton pair production.  The process of $J/\Psi$
    resonance production (with its decay into a lepton pair)
    $p + \bar p \to J/\Psi  + X \to l^{+}l^{-} + X$
    ($l = \mu, e$) is simulated by use of PYTHIA6.4 which includes
    a sizeable set of parton level subprocesses
    that can  give a contribution to this process.
    This process was chosen  to be a benchmark 
    process for PANDA experiment (see TPR 
    \cite{PANDATPR}). Thus, the obtained here
    kinematical distributions of the final 
    state leptons shall have a practical application. 
    In the following we shall call this process as
    {\it resonance production} to distinguish it from
    the process considered in Section 2, where the
    invariant mass of two leptons produced in quark
    level subprocess $q\bar{q} \rightarrow l^{+}l^{-}$
    has a {\it continuum} spectrum.

      Section 4 includes  the distributions 
    of the invariant mass and some other physical 
    variables which are characteristic for the signal 
    lepton pair as a whole system.
    The most interesting  among them is the total 
    transverse momentum of a lepton pair which is 
    connected with the intrinsic transverse velocity
    of a quark inside the proton. 

    In Section 5 we estimate  the size of the
    kinematical region in $x-Q^2$ plane which
    can be available for measuring the quark distributions
    in PANDA experiment.

    The problems connected  with the background from the fake 
    leptons, which may appear together  with the signal lepton pair
    in one and the same event due to meson decays,
    as well as  with  the background from other than
    $q\bar{q} \rightarrow l^{+}l^{-}$ subprocesses, are  discussed in 
    Sections  6 and 7, correspondingly.
   Also we present a set of cuts which
    allows to separate the background events from  
    the signal ones. The efficiencies of the 
    proposed cuts are also given.  
   
    In Section  8  we outline some important 
   physical measurements which can be done by studying
   the lepton pair production at the energies available 
   for PANDA experiment.

     It is worth  mentioning  that the results 
    presented here  can also be 
   useful  for the future physical analysis 
   of hadron decays  at   PANDA because  the contribution from
   $q\bar{q} \rightarrow l^{+}l^{-}$ events
   may be one of the main background sources
   in this kind of a study.


\section{Distributions of leptons produced  in $p\bar p$ collisions}

~~~    We use PYTHIA6  to generate  two samples 
   (separately for muons and electrons) of 100 000 
   "$\bar{p}p \rightarrow l^{+}l^{-} + X$" events
   which include the $2\rightarrow 2$  quark level 
   ${q\bar{q} \to \gamma^{*} \to l^{+}l^{-}}$ 
   subprocess. 
 In the following,  these events will be called as "signal events", and
   the muons/electrons produced in this  subprocess
   will  be called  as "signal" leptons. The fake
   leptons which are produced in hadron (mainly
   mesons) decays  in the same signal event will
   be called  "decay"  leptons. The simulation
   is done starting from the assumption  of the ideal
   muon system and electromagnetic  calorimeter
   covering  $360^{o}$.  No any cuts are used in 
   the following Sections. They will be discussed
   in  the subsection 6.5 and the Section 8. 

   We consider the case where both initial-state
   radiation (ISR) and final-state radiation 
   (FSR) are switched on simultaneously
   by choosing the corresponding  values of  PYTHIA parameters. 
   Also  we have used the CTEQ3L parametrization
   of parton distributions and the default value  of the parameter, 
   which allows one to   take into account the primordial 
   $k_T$-effect (the effect of quark Fermi  motion inside the hadron). 

     First we consider the distributions of some 
   physical variables which describe the kinematics
   of individual leptons  belonging to the
   $l^{+}l^{-}$ pair. Simulation shows that there 
   is no difference between $e^{+}e^{-}$ and
   $\mu^{+}\mu^{-}$  distributions. In all the 
   following figures
 the    vertical axis shows the number of events (per bin)
   that may be expected per  year  ($10^{7}sec$) for
   the luminosity  $L= 2 \cdot 10^{32} cm^{-2} s^{-1}$
    (=2 $\cdot 10^5 mb^{-1} s^{-1}$). 
   The total number of expected events  per year
   are shown as  "Integral" values in the figures.
   Let us underline that this number can be
   treated only as an estimation because it strongly
   depends on the models used in PYTHIA which
   is  basically designed for much higher energies.

  \begin{figure}[!ht]
     \begin{center}
\vskip -0.5cm       
   \mbox{a)\includegraphics[width=7.2cm, height=5.2cm]{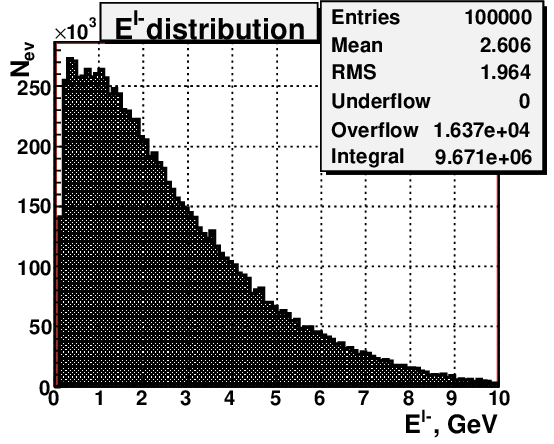}}  
   \mbox{b)\includegraphics[width=7.2cm, height=5.2cm]{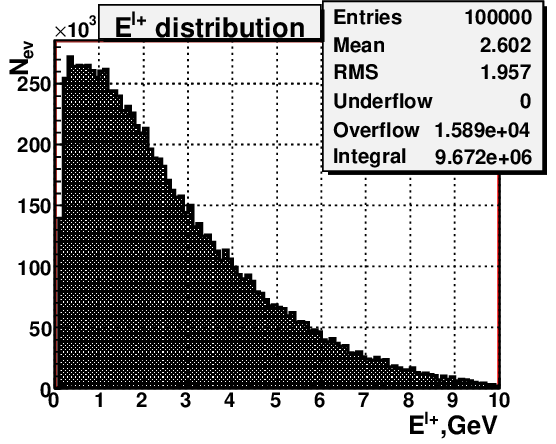}}  \\
   \mbox{c)\includegraphics[width=7.2cm, height=5.2cm]{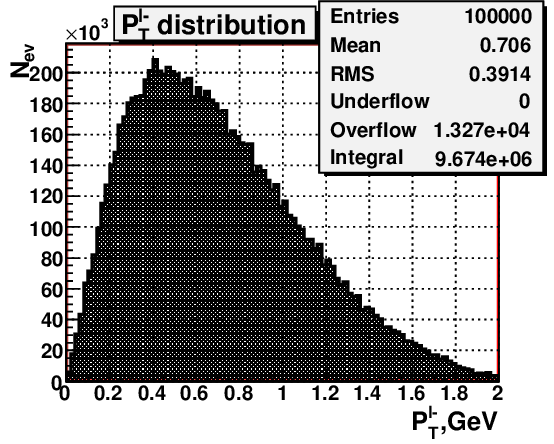}}  
   \mbox{d)\includegraphics[width=7.2cm, height=5.2cm]{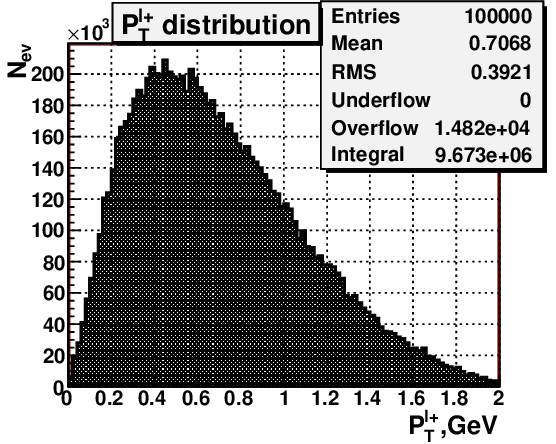}}  \\
   \mbox{e)\includegraphics[width=7.2cm, height=5.2cm]{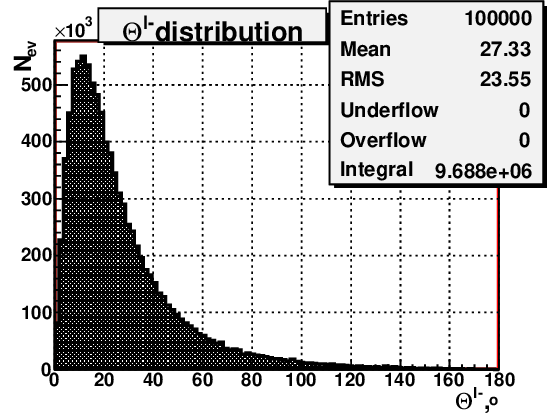}}  
   \mbox{f)\includegraphics[width=7.2cm, height=5.2cm]{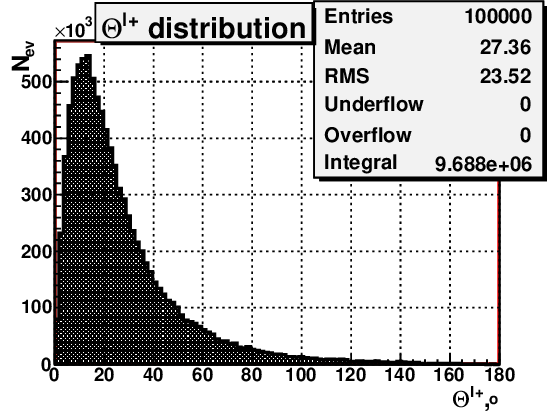}}  \\
		      		          
   \caption{\small \it Distributions of number of signal leptons versus:
 the  energy  $E^{l^{(+/-)}}$  (top row,  plots {\bf a} and {\bf b}), 
 the modulus of  transverse momentum  $P_{T}^{l^{(+/-)}}$
 (middle row,  plots {\bf c} and {\bf d})  and the  polar angle
  $\theta^{l^{(+/-)}}$  (bottom row,  plots {\bf e} and {\bf f}). 
     Left column:   $l^{-}$,  right column  $l^{+}$.}	 
     \end{center}    
\vskip -0.5cm           
     \end{figure}

   The distributions of the number of generated signal 
   events  ($N_{ev}$) versus the 
   the energy $E^{l^{(+/-)}}$ of the signal leptons,
   as well as versus  the modulus of the transverse
   momentum  $P_T^{l^{(+/-)}}$ and of the polar
   (zenith) angles $\theta^{l^{(+/-)}}$, measured
   from the z-axis directed along the beam line,
   are given (top to bottom) in  Fig.2. The left 
   column of Fig.2 is for $l^{-}$ distributions
   and the right one is for $l^{+}$. 
   One can see from the top row of Fig.2 (plots 
   {\bf a} and {\bf b}) that the most part of leptons
   energy is contained in the interval
   $0 < E^{l} < 10$ GeV. Its spectrum has a mean
   value $<E^{l}> = 2.6 $ GeV and a peak at
   $E^{l}_{peak} \approx 0.5$ GeV.  The $P_{T}^{l}$ 
   spectrum (see the  middle row plots
   {\bf c} and {\bf d} of Fig.2) has an analogous
   peak at $P^{l}_{T_{peak}} \approx 0.4$ GeV.
   Behind their peaks both $E^{l}$  and $P_T^{l}$
   spectra fall rather steeply. The main part of 
   $P_T^{l}$ spectrum is confined  within a
   rather  narrow interval $0 < P_T^{l} < 2$ GeV.

   The number of events  spectrum versus
   the polar angle  ${\theta^{l}}$ (see bottom 
   row plots {\bf e} and {\bf f} of Fig.2) has a
   peak around ${\theta^{l}} \approx 10^{o}$ and
   the mean value $ <\theta^{l}> = 27.3^{o}$. One
   sees that while the most of signal leptons fly
   in the forward direction ($\theta^{l} < 90^{o}$)
   there is still a small number of them which fly 
   into the back hemisphere ($\theta^{l} > 90^{o}$).

  \begin{figure}[!ht]
     \begin{center}
\vskip -0.5cm       
   \mbox{a)\includegraphics[width=7.2cm, height=5.2cm]{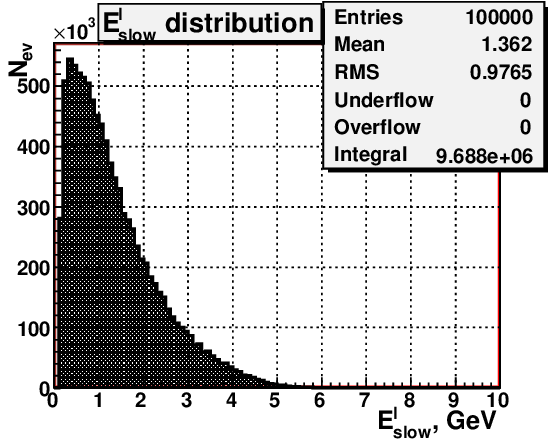}}  
   \mbox{b)\includegraphics[width=7.2cm, height=5.2cm]{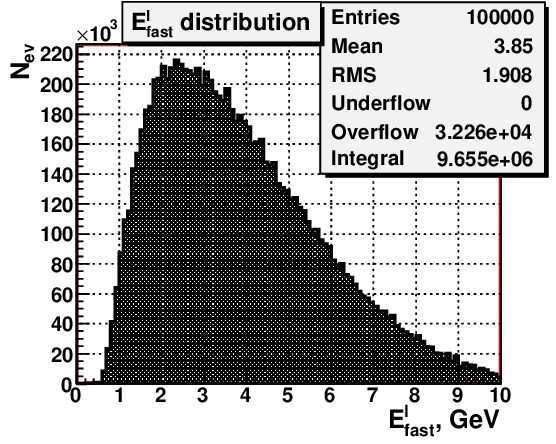}}  \\
   \mbox{c)\includegraphics[width=7.2cm, height=5.2cm]{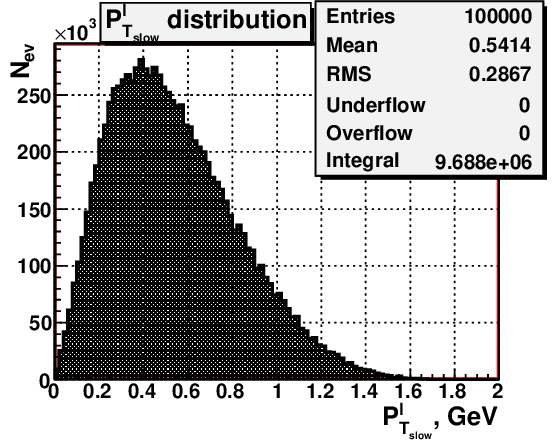}}  
   \mbox{d)\includegraphics[width=7.2cm, height=5.2cm]{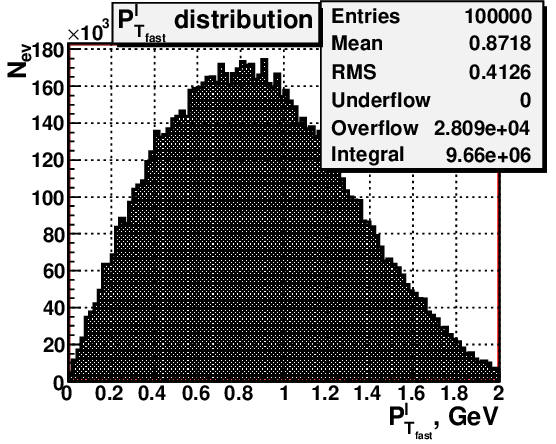}}  \\
   \mbox{e)\includegraphics[width=7.2cm, height=5.2cm]{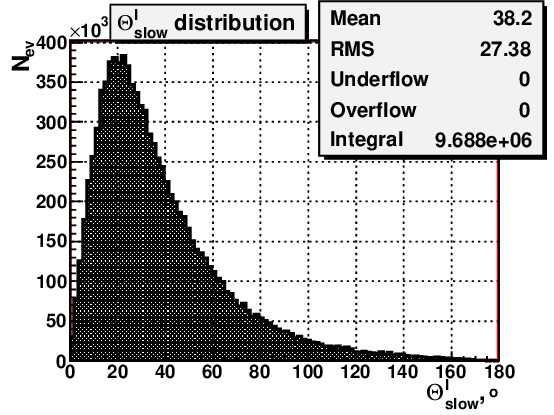}}  
   \mbox{f)\includegraphics[width=7.2cm, height=5.2cm]{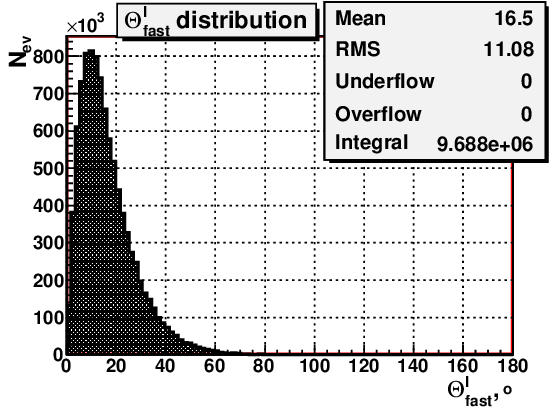}}  \\	
		      	          
 \caption{\small \it Distributions of number of signal leptons versus:
the  energy  $E^{l^{(+/-)}}$ (top row,  plots {\bf a} and {\bf b}), 
  the modulus of  the transverse momentum  $P_{T}^{l^{(+/-)}}$
(middle row,  plots {\bf c} and {\bf d})  and  the polar angle 
$\theta^{l^{(+/-)}}$ (bottom row,  plots {\bf e} and {\bf f}) .
  Left column:  low energy "slow" leptons, 
  right column:  high energy "fast" leptons.}	 
     \end{center}    
\vskip -0.5cm           
     \end{figure}
  
   Fig.3 includes the set  of plots done separately
  for the  signal leptons having the  largest energy
  $E^{l}_{fast}$ (right column) in the lepton pair 
  and for the leptons  having a smaller energy
  $E^{l}_{slow}$ in the pair (left column). We shall
  call  them, correspondingly, as "fast" and "slow"
  leptons. 
     One can see that the energy spectrum of  
   fast signal leptons (plot {\bf b}) 
   increases rather fast from the point  
   $E^{l}_{fast} \approx 0.5$ GeV 
   (more than  $90\%$ of fast leptons have
   $E^{l}_{fast} > 1 $ GeV) up to the peak  position  at the point
$E^{l}_{fast} \approx 2.5$ GeV ($ < E^{l}_{fast} >$ = 3.85 GeV). Then it
   smoothly vanishes at  $ E^{l}_{fast} = 10$ GeV.

     In contrast to this picture, the analogous
   spectrum of the  less energetic  signal
   leptons (plot {\bf a}) starts sharply
   from zero and  reaches  a peak at
   $E^{l}_{slow} \approx 0.4$ GeV 
   (where the spectrum of the fast leptons only
   starts). Then it goes down and practically vanishes at
   the point $E^{l}_{slow} \approx 5$ GeV. 
    One may see that the  energy
   spectrum of  slow leptons  plot {\bf a} in a pair is  
   more than by half shorter  than  that one
   of  fast leptons plot {\bf  b} and the mean value of 
   slow leptons energy $ < E^{l}_{slow} > = 1.36$ GeV is
   about 3 times less than the mean energy
   of   fast leptons $ < E^{l}_{fast} > = 3.85$ GeV.

    The difference between  the  $PT^{l}$ 
   spectra  of fast and slow leptons is not so large (see, 
   correspondingly, plots {\bf d} and {\bf c}
   in the middle row of Fig.3).  They
   differ only by about 400 MeV shift to the 
   left of the peak position of the slow leptons
   spectrum and about 340 MeV analogous shift
   of their mean transverse momentum value.   
   Both  of these spectra  demonstrate that the main part of slow 
   and fast leptons has  $PT^{l} > 0.2$ GeV.

    The bottom row of the  plots demonstrates  
   that  the polar (zenith) angle 
spectrum of less energetic leptons ${\theta^{l}_{slow}}$ (Fig.3 {\bf e})
   is shifted to the higher values as compared to the spectrum of  fast 
   leptons ${\theta^{l}_{fast}}$ (Fig.3 {\bf f}).
   Its mean value  $ < {\theta^{l}_{slow}} > = 38.2^{o}$
   is more than twice  as large as the 
   analogous mean value of fast leptons: 
   $ < {\theta^{l}_{fast}} > = 16.5^{o}$.
   Thus, we  can conclude that  almost 
   all fast leptons fly in the forward direction
   (${\theta^{l}_{fast}} < 80^{o}$) and their
  spectrum practically finishes at ${\theta^{l}_{fast}} \approx 60^{o}$ 
 (plot {\bf f}) while about  $17\%$ of slow leptons (plot {\bf e}) have 
   ${\theta^{l}_{slow}} > 60^{o}$.  It is worth 
   noting  that  about   $5\%$ of slow leptons
   may scatter into the back hemisphere.

\begin{figure}[!ht]
     \begin{center}
\vskip -0.5cm       
   \mbox{a)\includegraphics[width=7.2cm, height=5.2cm]{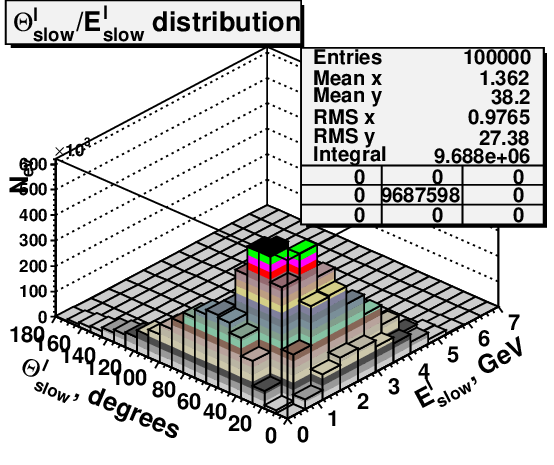}}  
   \mbox{b)\includegraphics[width=7.2cm, height=5.2cm]{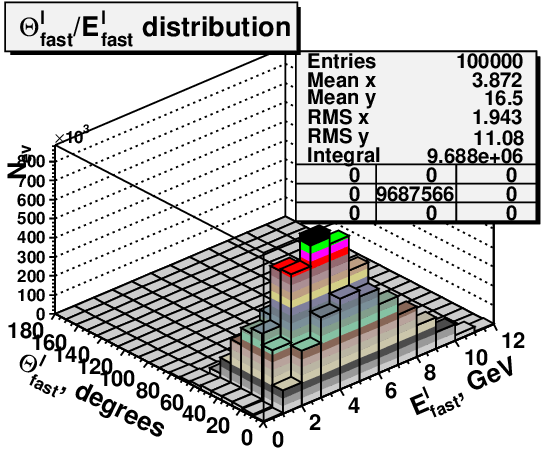}}  \\	         
   \mbox{c)\includegraphics[width=7.2cm, height=5.2cm]{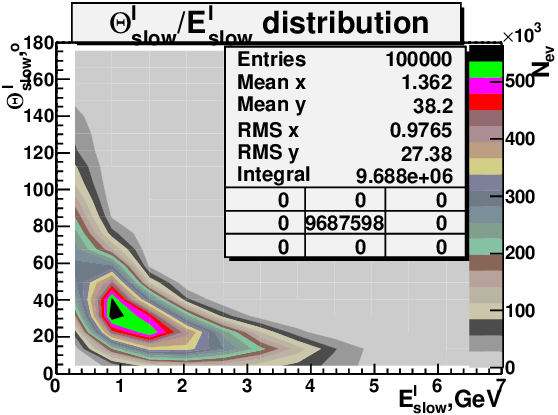}}  
   \mbox{d)\includegraphics[width=7.2cm, height=5.2cm]{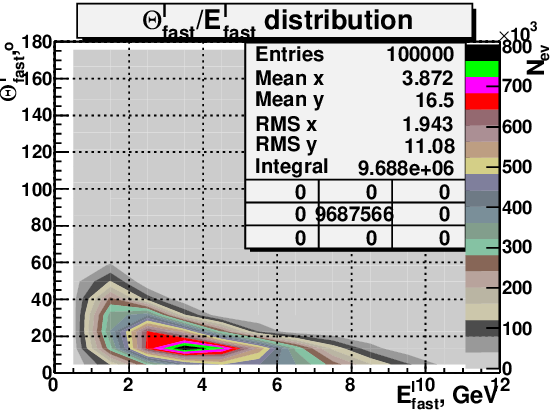}}  \\	          
     \caption{\small \it Angle-Energy correlations.
 {\bf a)} and  {\bf c): }${\theta^{l}_{slow}}/{E^{l}_{slow}}$ 
                           correlation for slow leptons, 
{\bf b)} and  {\bf d): }${\theta^{l}_{fast}}/{E^{l}_{fast}}$
		           correlation for fast leptons.
                Plots  {\bf c)} and {\bf d)} are the 
                pojections of 3D plots  {\bf a)} and  
                {\bf b)} on ${\theta^{l}}-{E^{l}}$ planes.}
 \end{center}    
\vskip -0.5cm           
     \end{figure}
 
    Fig.4  contains two 3-Dimensional "Angle-Energy"
    correlation plots for slow  ${\theta^{l}_{slow}}/{E^{l}_{slow}}$
    (plot $\bf{a }$) and fast    ${\theta^{l}_{fast}}/{E^{l}_{fast}}$
    (plot $\bf{b }$) leptons in signal pairs.
     Their vertical axes show the distribution of 
    number of signal events ($N_{ev}$)   multiplied by $10^{3}$. 
    2-dimensional   plots $\bf{c }$ and $ \bf{ d}$   are
    obtained by projection  of plots $\bf{a }$ 
    and  $ \bf{b}$ onto the $\theta-E$ planes.
    This allows to show the boundary  contours
    of regions with different density of number
    of events.  The right-hand color vertical strip
    in each plot $\bf{c }$ and  $ \bf{ d}$ shows 
    the correspondence between the  contour color
    and the density of events. This color strip
    plays the role of the vertical z axis with the
    number of events $N_{ev}$ shown in plots {\bf a}
    and {\bf b}. Plots {\bf c} and {\bf d} of Fig.4  
    show the  kinematical regions in $\theta-E$ 
    plane which are covered by slow and fast leptons,
    respectively. Plot {\bf c}   of Fig.4 demonstrates that  
    the value of the polar angle of slow  leptons  
    ${\theta^{l}_{slow}}$ drops much steeply with the 
    growth of their energy as compared to the
    behavior of the polar angle of fast lepton
    ${\theta^{l}_{fast}}$ shown at the plot $\bf{d }$.

 After discussion  of individual lepton distributions let us turn to the
 distributions that characterize the produced pair of leptons as
    a whole system. Fig.5 shows the Energy-Energy  
    $E^{l}_{slow}/E^{l}_{fast}$ (plot{\bf s} {\bf a} and {\bf c})
    and Angle-Angle   ${\theta^{l}_{slow}}/{\theta^{l}_{fast}}$ 
    (plots {\bf b}  and {\bf d}) correlations.
    Plots {\bf c} and {\bf  d}  are the projections of 3D plots
    {\bf a} and  {\bf b} on  $E^{l}_{slow}-E^{l}_{fast}$
    and ${\theta^{slow}}-{\theta^{fast}}$   planes, correspondingly.

   \begin{figure}[!ht]
     \begin{center}
\vskip -0.5cm       
 
   \mbox{a)\includegraphics[width=7.2cm, height=5.2cm]{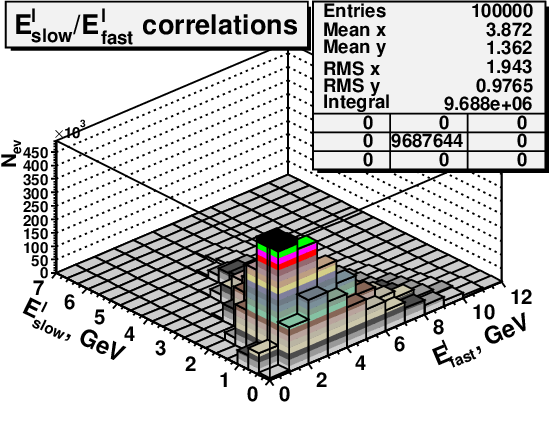}}  
   \mbox{b)\includegraphics[width=7.2cm, height=5.2cm]{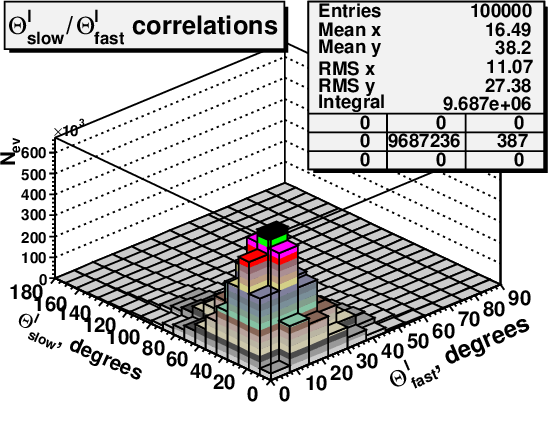}}  \\	          
   \mbox{c)\includegraphics[width=7.2cm, height=5.2cm]{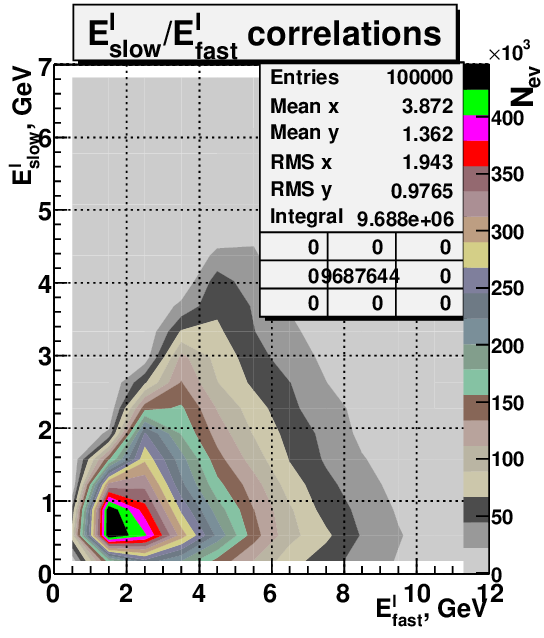}}  
   \mbox{d)\includegraphics[width=7.2cm, height=5.2cm]{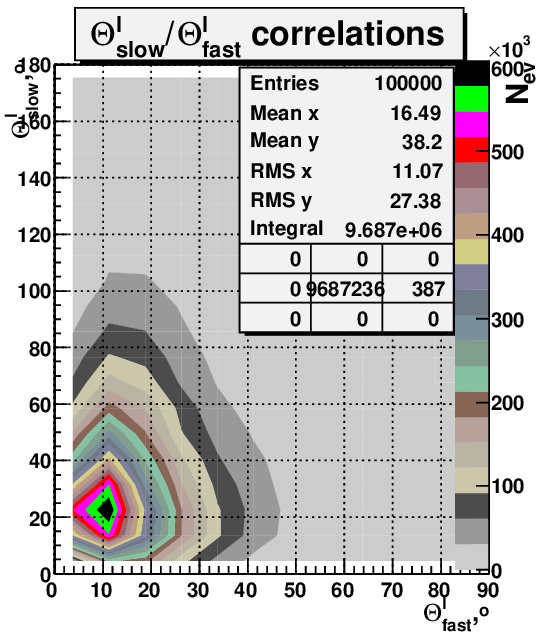}}  \\	          
     \caption {\small \it
 {\bf a)} and {\bf c):}
 Energy-Energy $E^{l}_{slow}/E^{l}_{fast}$ correlations. 
{\bf b)} and {\bf d):}
 Angle-Angle $\theta^{l}_{slow}/\theta^{l}_{fast}$ correlations.
 Plots {\bf c)} and {\bf  d)}  are the projections of 3D plots
 {\bf a)} and  {\bf b)} on  $E^{l}_{slow}-E^{l}_{fast}$
 and  ${\theta^{slow}}-{\theta^{fast}}$  planes  correspondingly.}	 
 \end{center}    
\vskip -0.5cm           
 \end{figure}

Data taking of searched signal events is 
strongly influenced by the  cuts on lepton energies which can be
     imposed from below to suppress  the electronic
     noise  and  some  other background 
     which can be provided by detector effects. 
      The analysis of  the plots {\bf a} and {\bf c} of Fig.5 is 
     summarized in the Table 1. It demonstrates 
     (in $\%$) the loss of signal events
     due to  application of the kinematical cut 
     $ E_{fast}^{l}, E_{slow}^{l} \geq E_{cut}$
     which sets the lower  limit  $E_{cut}$ on the
     value of lepton energy.   
 
\begin{table}[!hbt]
\vskip -0.5cm        
\caption{Efficiency of the  $ E_{fast}^{l}, E_{slow}^{l} \geq E_{cut}$ cut}
\begin{center}
\begin{tabular}{|c|c|}              \hline 
 $ E_{cut}$ ( in GeV) & The loss of signal events ( in $\%$)  \\\hline
 0.5  &     24 \\\hline  
 1.0  &     45 \\\hline     
\end{tabular}
\end{center}
\end{table}      
  
    The efficiency of collection of signal events
    which contain  leptonic $l^{+}l^{-}$ pairs
    also depends on the angle coverage by the muon 
    system and the  Electromagnetic Calorimeter (ECAL).
    Plots  $\bf{b}$ and  $\bf{d}$ of Fig.5 show the
    Angle-Angle  ${\theta^{l}_{slow}}/{\theta^{l}_{fast}}$
    lepton correlation.   The results of its analysis
    are given in the Table 2 which demonstrates
    which part  (in $\%$) of signal events would be
    lost due to imposing  the upper limit  $\theta_{cut}$ 
(i.e. ${\theta^{l}_{slow}}, {\theta^{l}_{fast}} \leq \theta_{cut}$  cut)  on  the size of muon  system or  ECAL.

\begin{table}[!hbt]
\caption{Efficiency of the 
        $ {\theta^{l}_{slow}}, {\theta^{l}_{fast}} \leq
	 \theta_{cut}$  cut}
\begin{center}
\begin{tabular}{|c|c|}              \hline 
$ \theta_{cut}^{0}$ ~( in~$^0$)    & 
 The loss  of signal events ( in $\%$ )  \\\hline
 20  &     80 \\\hline  
 40  &     39 \\\hline  
 60  &     17 \\\hline  
 90  &     5 \\\hline   
\end{tabular}
\end{center}
\vskip -0.5cm
\end{table} 

  The last line of Table 2 shows  that even in the
  case when the muon system or the  ECAL  would  cover the 
  angle region  $\theta^{l} \leq 90^{o}$, about  $5\%$ of the events
  containing the  $l^{+}l^{-}$ signal pairs  would be lost.
 Nevertheless, such geometrical boundary allows to keep about 95$\%$ of
  signal events with electron or muon pairs. 
  Therefore, we consider this choise of
  polar angle upper limit as a preferable one for 
  the study of MMTDY process of lepton pair production
 (with the continuous  mass spectrum of this pair). \\


    \section{ Leptons from J/$\Psi$ decay} 


   ~~~ The process  of J/$\Psi$ resonance production
    with its  further decay into a lepton ($l=\mu, e$)
    pair $p + \bar p \to J/\Psi  + X \to l^{+}l^{-} + X$
(see one of the possible reaction on Fig.6) was considered in the PANDA 
    TDR \cite{PANDATPR} as one of the benchmark processes. Therefore, modeling of the kinematical
    (energy, transverse momentum and angle) distributions
    of final state leptons is of practical interest.

   \begin{figure}[!ht]
     \begin{center}
    \mbox{\includegraphics[width=12cm, height=7.0cm]{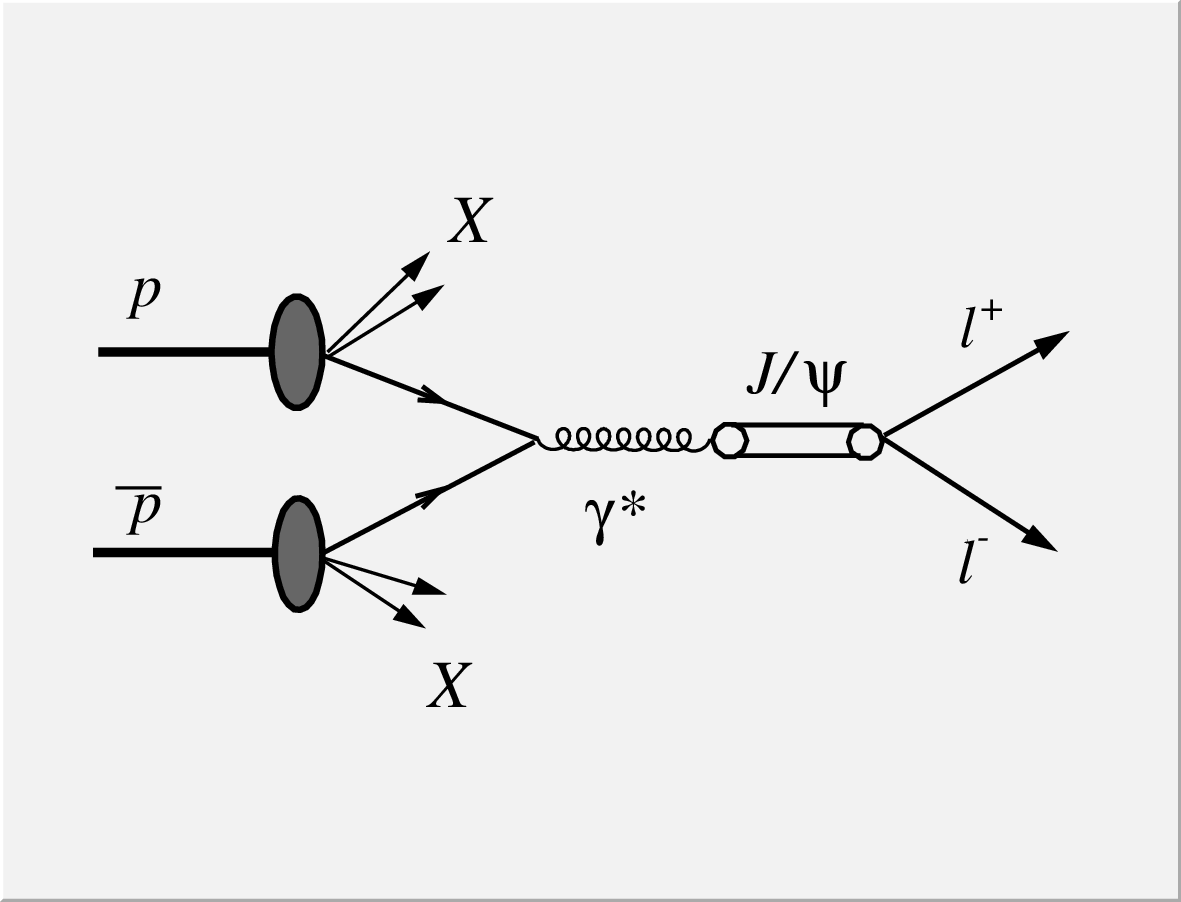}}
 \caption{\small \it $\bar{p}p \rightarrow J/\Psi  + X \to
     l^{+}l^{-} + X$ process }
     \end{center}
     \end{figure}

   Like in previous Section, we use  the same event
   generator PYTHIA6.4 which includes  the following
   set of subprocesses with J/$\Psi$  production:\\ \\
  1) $q_{i} \bar q _{i} \to \gamma^{*} \to c\bar c \to
   J/\Psi \to l^{+}l^{-} +  X$
~~~~~~~~~~~
~~~~ 86)  $g g \to J/\Psi  + g \to l^{+}l^{-} + X$ \cite{R.Baier} \\
106)  $g g \to J/\Psi  + \gamma \to l^{+}l^{-} + X$ \cite{M.Drees} 
~~~~~~~~~~~~~~~
~~~~~ 421) $  g g   \to  c\bar c  [^3S_1^{(1)} ]  g  \to  l^{+}l^{-} + X$ \cite{G.T.Badwin} \\
422) $  g g   \to  c\bar c  [^3S_ 1^{(8)} ]  g  \to   l^{+}l^{-} + X$ \cite{G.T.Badwin}
~~~~~~~~~~~~~~~~~~~~
423) $  g g   \to  c\bar c  [^3S_ 0^{(8)} ]  g  \to  l^{+}l^{-} + X$ \cite{G.T.Badwin} \\
424) $  g g   \to  c\bar c  [^3P_ J^{(8)} ]  g  \to  l^{+}l^{-} + X$ \cite{G.T.Badwin}
~~~~~~~~~~~~~~~~~~~~
425) $  g q   \to  c\bar c  [^3S_ 1^{(8)} ]  q  \to  l^{+}l^{-} + X$ \cite{G.T.Badwin}\\
426) $  g q   \to  c\bar c  [^3P_ J^{(8)} ]  q  \to  l^{+}l^{-} + X$ \cite{G.T.Badwin}
~~~~~~~~~~~~~~~~~~~~
427) $  g g   \to  c\bar c  [^3S_ 1^{(1)} ]  q  \to  l^{+}l^{-} + X$ \cite{G.T.Badwin}\\ 
428) $  q \bar q \to  c\bar c  [^3S_ 1^{(8)} ]  g  \to  l^{+}l^{-} + X$ \cite{G.T.Badwin}
~~~~~~~~~~~~~~~~~~~~
429) $  q \bar q \to  c\bar c  [^1S_ 0^{(8)} ]  g  \to  l^{+}l^{-} + X$ \cite{G.T.Badwin}\\
430) $  q \bar q \to  c\bar c  [^3P_ J^{(8)} ]  g  \to  l^{+}l^{-} + X$ \cite{G.T.Badwin}
~~~~~~~~~~~~~~~~~~
431) $  g g    \to  c\bar c  [^3P_0^{(1)} ]  g  \to  l^{+}l^{-} + X$  \cite{G.T.Badwin}\\
432) $  g g    \to  c\bar c  [^3P_1^{(1)} ]  g  \to  l^{+}l^{-} + X$ \cite{G.T.Badwin}
~~~~~~~~~~~~~~~~~~~
433) $  g g    \to  c\bar c  [^3P_2^{(1)} ]  g  \to  l^{+}l^{-} + X$ \cite{G.T.Badwin}\\ 
434) $  g q    \to  c\bar c  [^3P_0^{(1)} ]  q \to   l^{+}l^{-} + X$ \cite{G.T.Badwin}
~~~~~~~~~~~~~~~~~~~~
435) $  g q    \to  c\bar c  [^3P_1^{(1)} ]  q  \to  l^{+}l^{-} + X$ \cite{G.T.Badwin} \\ 
436) $  g q   \to  c\bar c  [^3P_2^{(1)} ]  q  \to   l^{+}l^{-} + X$ \cite{G.T.Badwin}
~~~~~~~~~~~~~~~~~~~~
437) $  q q   \to  c\bar c  [^3P_0^{(1)} ]  g  \to   l^{+}l^{-} + X$ \cite{G.T.Badwin}\\ 
438) $  q \bar q \to  c\bar c  [^3P_1^{(1)} ]  g  \to   l^{+}l^{-} + X$ \cite{G.T.Badwin}
~~~~~~~~~~~~~~~~~~~~
439) $  q \bar q \to  c\bar c  [^3P_2^{(1)} ]  g  \to   l^{+}l^{-} + X$ 
\cite{G.T.Badwin}\\

 The main contribution to the total cross section of J/$\Psi$ production 
 \footnote { Let us note that different theoretical
   models predict different values of J/$\Psi$ production cross sections. }
  comes from the following three subprocesses:\\
  1)~~~ $q_{i} \bar q _{i} \to \gamma^{*} \to c\bar c \to
   J/\Psi \to l^{+}l^{-} +  X$\\
428) $  q \bar q \to  c\bar c  [^3S_ 1^{(8)} ]  g  \to   l^{+}l^{-} + X$ \\
430) $  q \bar q \to  c\bar c  [^3P_ J^{(8)} ]  g   \to   l^{+}l^{-} + X$ \\


  \begin{figure}[!ht]
     \begin{center}
\vskip -0.5cm       
   \mbox{a)\includegraphics[width=7.2cm, height=5.2cm]{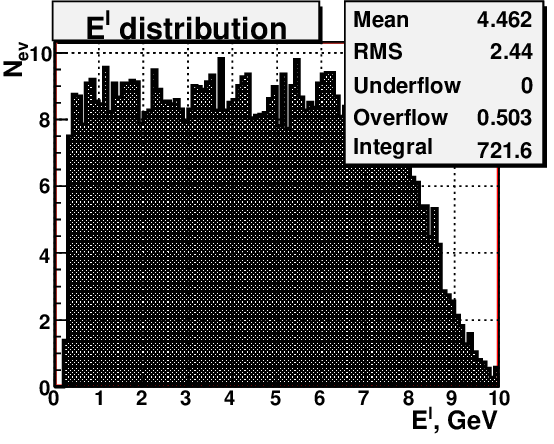}}  
   \mbox{b)\includegraphics[width=7.2cm, height=5.2cm]{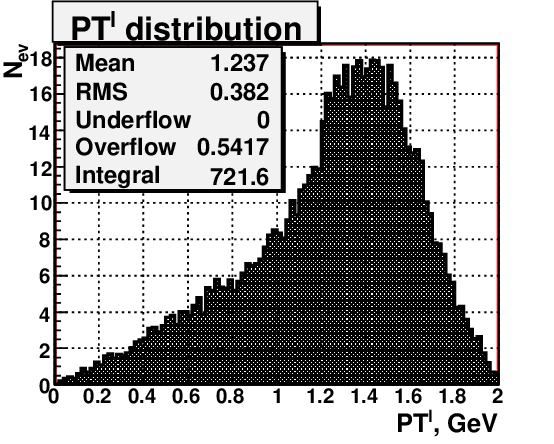}}  \\
   \mbox{c)\includegraphics[width=7.2cm, height=5.2cm]{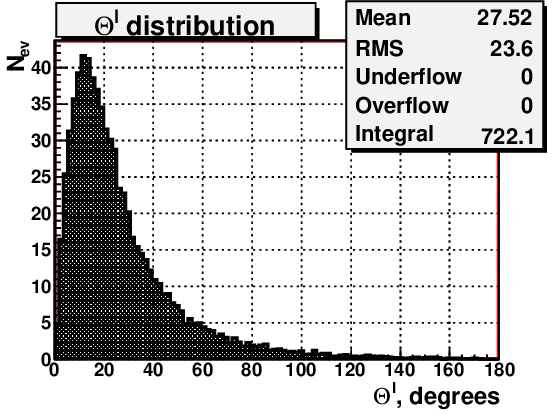}}  
   \mbox{d)\includegraphics[width=7.2cm, height=5.2cm]{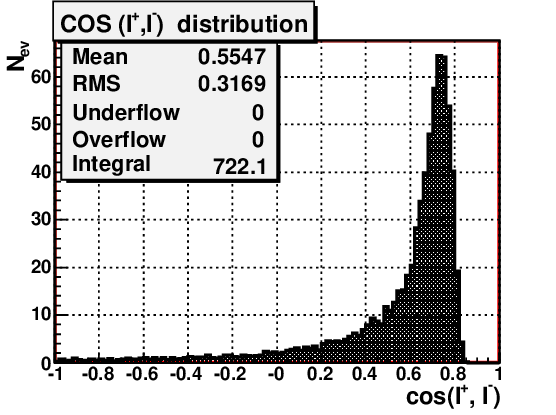}}  \\
		      		          
     \caption{\small \it Distributions of the leptons
                       from $J/\Psi$ decay into $ l^{+}l^{-} +  X$:
   {\bf a)}  energy  $E^{l^{(+/-)}}$, 
   {\bf b)}  modulus of    the transverse momentum  $P_{T}^{l^{(+/-)}}$;
    {\bf c)}   polar angle $\theta^{l^{(+/-)}}$;
     {\bf d)}  cos of  the  opening angle between the leptons
     in $l^{+}l^{-}$ pair.}	 
     \end{center}    
\vskip -0.5cm           
     \end{figure}

 Distributions of the final state leptons, produced 
   in J$/\Psi$ decay,  are shown in Fig.7.
  They are obtained  without a use of any cuts and
  cover the same ranges as the leptons produced
  in the continuum case (see Fig.2). From comparison
  of these two Figures one can see that the 
  energy and momentum distributions presented 
  in Fig.7 look very different to those of Fig.2.
  Thus, plot {\bf a}  of Fig.7 shows a
  rather flat distribution of the number of events
  versus the energy of leptons from J$/\Psi$ decay,
  while the analogous plots {\bf a} and  {\bf b}
  of Fig.2 demonstrate that the most part of
  leptons produced in the continuum case have
  small energies $E^{l} < 1$ GeV. Analogously,
  as seen from the plot {\bf b} of Fig.7,
  the peak of $P_{T}^{l}$ distribution in a 
  case of J$/\Psi$ production   appears at the point
  $P_{T}^{l} \approx 1.4$ which is more than
  three times higher than the maximum value of $P_{T}^{l}$
  in  the plot {\bf b} of Fig.2 ($\approx 0.4$ GeV).
  As one can see from the plot {\bf d } of Fig.7,
  the maximum of cos $(l^{+},l^{-})$ is at $\approx$ 0.75.
  This value corresponds to the opening 
  angle between the leptons of $\approx 42^o$.

  \begin{figure}[!ht]
     \begin{center}
   \mbox{a)\includegraphics[width=7.2cm, height=5.2cm]{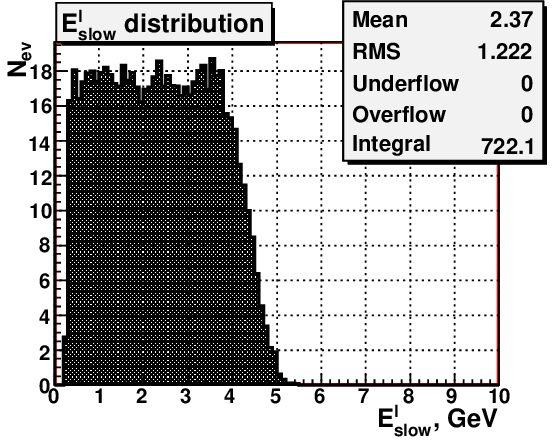}}  
   \mbox{b)\includegraphics[width=7.2cm, height=5.2cm]{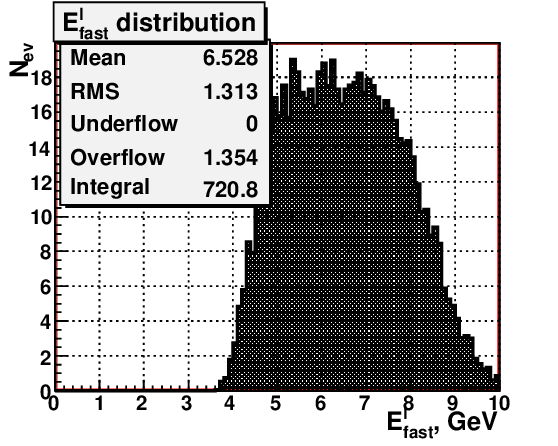}}  \\
   \mbox{c)\includegraphics[width=7.2cm,
 height=5.2cm]{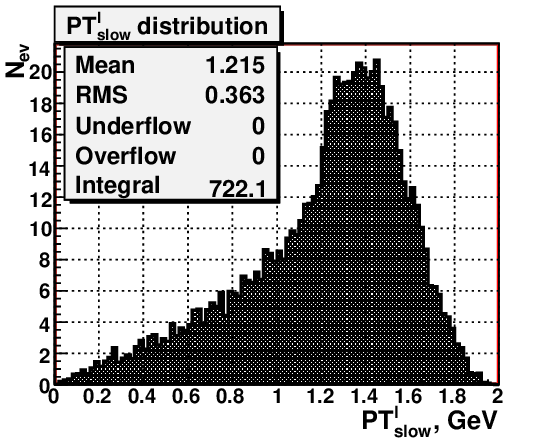}}  
   \mbox{d)\includegraphics[width=7.2cm, 
height=5.2cm]{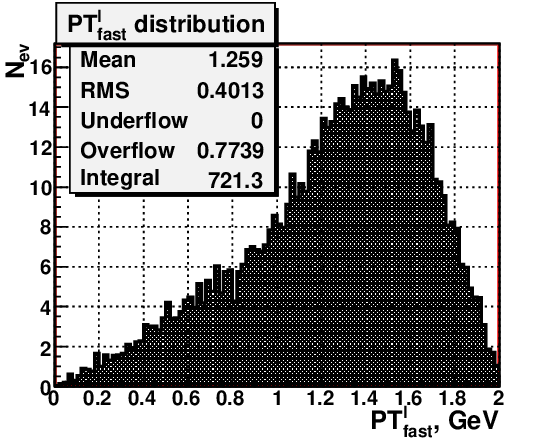}}  \\
   \mbox{e)\includegraphics[width=7.2cm, height=5.2cm]{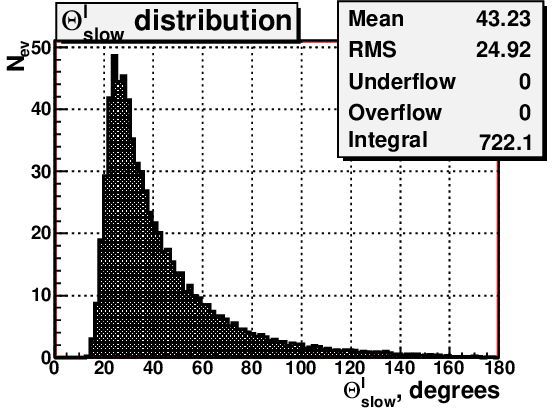}}  
   \mbox{f)\includegraphics[width=7.2cm, height=5.2cm]{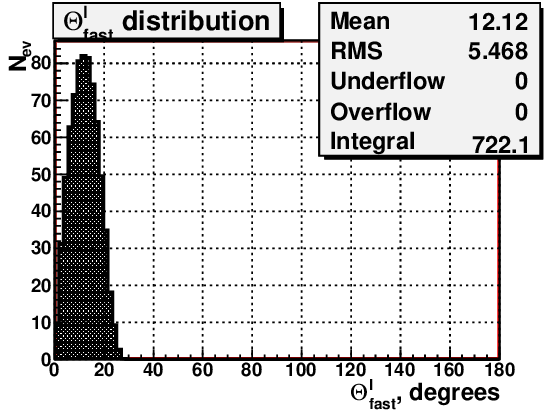}}  \\	
		      	          
     \caption{\small \it Distributions of the leptons  
       from $J/\Psi$ decay into $ l^{+}l^{-} +  X$:
       energy  $E^{l^{(+/-)}}$ (top row),  modulus of 
       the transverse momentum  $P_{T}^{l^{(+/-)}}$
       (middle row) and   polar angle $\theta^{l^{(+/-)}}$
       (bottom row). Left column:  low energy "slow" leptons, 
       right column:  high energy "fast" leptons.}	 
      \end{center}    
\vskip -0.5cm           
  \end{figure}
    
   Fig.8  is the analog of Fig.3. It presents the
   distributions of the same leptons produced in 
   J$/\Psi$ decay, but separately for "slow" and "fast" ones.
   One can see that the plots {\bf a} and {\bf b} of Fig.8
   look much more different from the  plots {\bf a} and 
   {\bf b} of Fig.3. The main difference is that the
   energy spectra of slow ($E_{slow}$) and fast ($E_{fast}$) 
   leptons, shown in the 
   plots {\bf a} and {\bf b} of Fig.8, cover very different
   energy intervals which have very small area of their
   overlapping in the region of $E^{l} \approx 4-5$ GeV. 
   It is also seen from the plots {\bf c} and {\bf d} of Fig.8
   that transverse momenta of both slow and fast leptons have
   a rather close mean values about   $P_{T}^{l} \approx 1.4$.
   Plots {\bf e} and {\bf f} of Fig.8 demonstrate a good
   separation of polar angle distributions of slow and fast 
   leptons. It is seen that the most part of the spectrum 
   of slow leptons lay behind the 20$^{o}$, while the 
   spectrum of fast leptons polar angles  ranges in the
   interval from 0$^{o}$ to 25$^{o}$ and has the mean value
   $\theta_{fast}^{l}=$12.12 GeV.

 \begin{figure}[!ht]
     \begin{center}
   \mbox{a)\includegraphics[width=7.2cm, height=5.2cm]{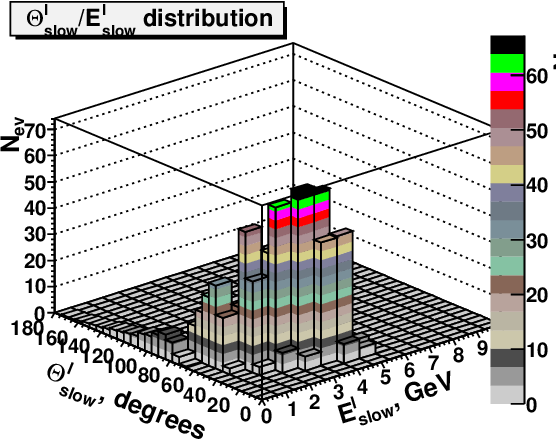}}  
   \mbox{b)\includegraphics[width=7.2cm, height=5.2cm]{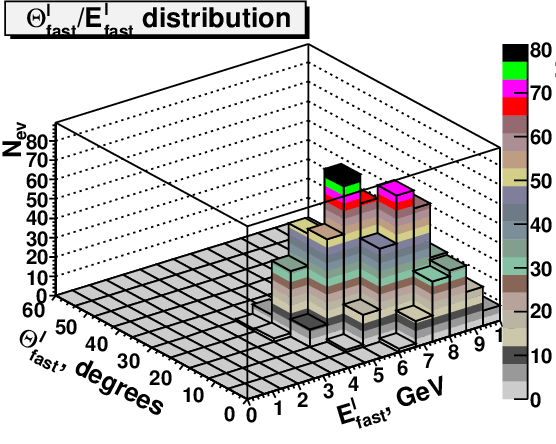}}  \\	          
      
   \mbox{c)\includegraphics[width=7.2cm, height=5.2cm]{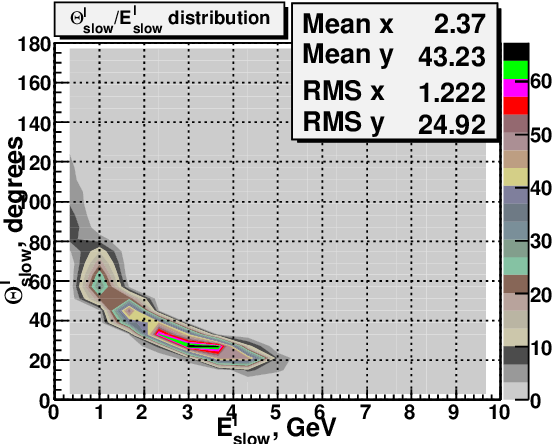}}  
   \mbox{d)\includegraphics[width=7.2cm, height=5.2cm]{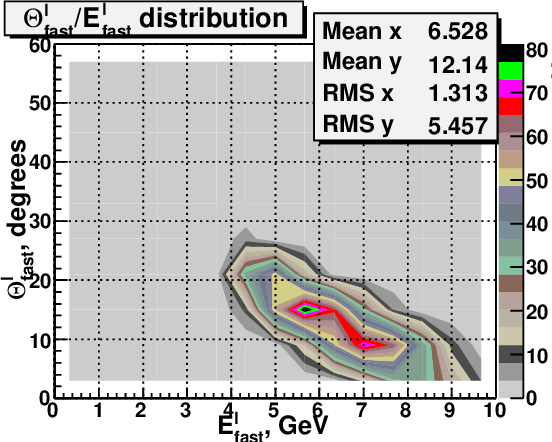}}  \\	          
     \caption{\small \it Angle-Energy correlations.
	       {\bf a)} and  {\bf c): }${\theta^{l}_{slow}}/{E^{l}_{slow}}$ 
                           correlation for slow leptons, 
	       {\bf b)} and  {\bf d): }${\theta^{l}_{fast}}/{E^{l}_{fast}}$
		           correlation for fast leptons.
Plots  {\bf c)} and {\bf d)} are the projections of 3D plots 
 {\bf a)} and  {\bf b)} onto ${\theta^{l}}/{E^{l}}$ plane. }

 \end{center}    
\vskip -0.5cm           
     \end{figure}
 
    Fig.9 includes the plots which show  Angle-Energy correlations
   of the leptons produced in $J/\Psi$ decay. Plots {\bf a} and {\bf c}
   are for slow leptons, plots {\bf b} and {\bf d} are for fast leptons.
   They are also very different from those shown in Fig.4 for the case
   of continuum lepton pair production via the process  
   $p\bar p + p \to \gamma^{*} \to l^{+}l^{-} + X$. It is seen that
   fast and slow leptons produced in $J/\Psi$ decay cover very
   different and well separated regions 
   (see plots {\bf c} and {\bf d} of Fig.9)  while in the continuum case
   (see plots {\bf c} and {\bf d} of Fig.4) the distributions of 
fast and slow leptons have  a wide area of their overlapping.
   These regions are rather narrow and have a strip form, that looks
   different from the case of continuum lepton pair production
   (see plots {\bf c} and {\bf d} of Fig.4).

     It is of interest to compare   the Energy-Energy
    and Angle-Angle correlation plots which are presented 
in Fig.10 for the case of lepton pair production from $J/\Psi$ resonance decay  and those shown in Fig.5 for leptons production in continuum  case. 
     It is seen from the plot {\bf c} of Fig.10 that the energy
     region covered in $E_{slow}-E_{fast}$-plane in the case
     of  $J/\Psi$-resonance production process fits well into
     the right corner of the region covered covered in the
     $E_{slow}-E_{fast}$-plane shown in the plot {\bf c}
     of  Fig.5 for the case of continuum mass MMTDY process
     of  lepton-antilepton pair production.

    \begin{figure}[!ht]
     \begin{center}
    
    \mbox{a)\includegraphics[width=7.2cm, height=5.2cm]{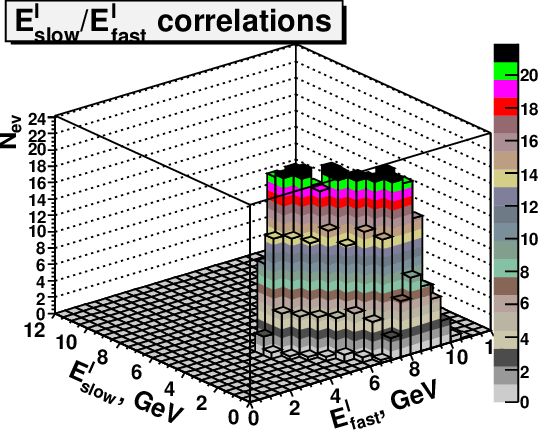}}  
   \mbox{b)\includegraphics[width=7.2cm, height=5.2cm]{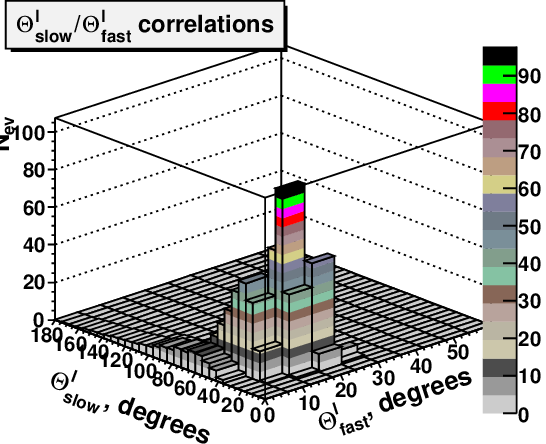}}  \\	          
   \mbox{c)\includegraphics[width=7.2cm, height=5.2cm]
{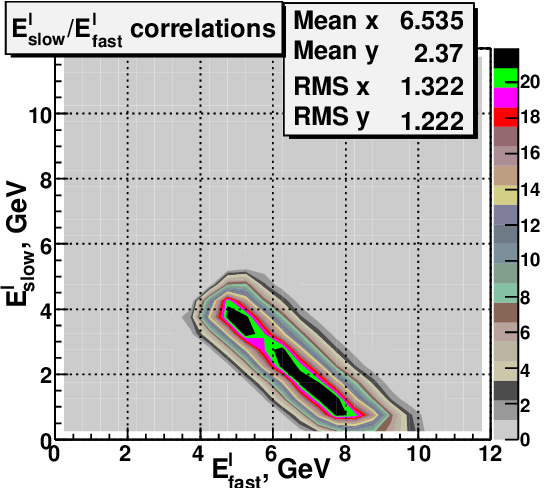}}  
   \mbox{d)\includegraphics[width=7.2cm, height=5.2cm]{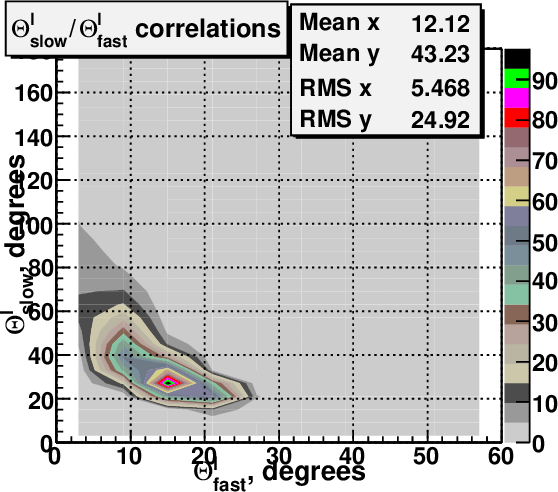}}  \\	          
     \caption {\small \it
 {\bf a)} and {\bf c):}
 Energy-Energy $E^{l}_{slow}/E^{l}_{fast}$ correlations, 
{\bf b)} and {\bf d):}
 Angle-Angle $\theta^{l}_{slow}/\theta^{l}_{fast}$ correlations.
 Plots {\bf c)} and {\bf  d)}  are the pojections of 3D plots
 {\bf a)} and  {\bf b)} onto  $E^{l}_{slow}/E^{l}_{fast}$
 and ${\theta^{slow}}-{\theta^{fast}}$  plane correspondingly. }	
    \end{center}    
\vskip -0.5cm           
     \end{figure}  

       Moreovere, the  analogous comparison of the 
    plot {\bf d} of Fig.10 with the plot
    {\bf d} of Fig.5 allows to make an important observation that
    the area covered in  the $\theta_{slow} - \theta_{fast}$ plane
    fits well into the analogous aria of Angle-Angle correlation
    plot in Fig.5. From here one can conclude that the choice of
    polar angle  boundary for muon system geometry, which was
    proposed in Section 2 basing on the analysis of MMTDY process, 
    would be quite suitable for the study of  both benchmark 
    $J/\psi$ and MMTDY channels of lepton-antilepton
    pair production.

    \section{Distributions of the invariant mass, energy \\
             and transverse momenta of lepton pairs}

   ~~~We consider here a set of physical 
   variables  which characterize a produced 
   lepton pair as a whole system.  
   These variables are constructed from the
   components  of the total  4-momentum of 
   initial state quark-antiquark  system 
   $P^{q\bar{q}}_{\alpha} =  P^{q}_{\alpha} + P^{\bar{q}}_{\alpha}$,
    ($\alpha=0,1,2,3$)       and its  analog 
   ${P^{l^{+}l^{-}}_{\alpha}=  P^{l^{+}}_{\alpha} + P^{l^{-}}_{\alpha}}$
   for a lepton  pair  ($P^{l^{\pm}}$ is the 4-momentum of lepton {\it l})
    \footnote{  $P^l=(P^{l}_{o}, {\bf P}^{l})$, 
   where ${\bf P}= (P_{x}, P_{y}, P_{z})$ 
   and $P_{o}= \sqrt{M^{2} + {\bf P^{2}}}$.}.
    Figs.11{\bf {a}}, 11{\bf {b}} show, correspondingly, the 
   distributions of the invariant masses
   of initial-state quark-antiquark pair   
\begin{equation}
 M_{inv}^{ q\bar{q} }=\sqrt( P^{q\bar{q}})^2,   
\end{equation}
    and the invariant mass of the final-state  lepton-antilepton pair 
\begin{equation}
  M_{inv}^{l^{+}l^{-}}=\sqrt( P^{l^{+}l^{-}} )^2=Q,
 ~~~{Q^2=q^2=( P^{l^{+}}_{\alpha} + P^{l^{-}}_{\alpha})^2 },
\end{equation} 
 been produced in the signal process $\bar{p}p \rightarrow l^{+}l^{-} + X$
    which goes through the quark level subprocess
    ${q\bar{q} \to \gamma^{*} \to l^{+}l^{-}}$.
   Both invariant mass distributions
   look rather similar.
   They are rather short and  drop  steeply with
   the growth of  invariant mass.
   The distribution of the invariant mass  $ M_{inv}^{ q\bar{q} }$ of the
   initial-state $q\bar{q}$-system   sharply starts at the point  
   $ M_{inv}^{ q\bar{q} } = 1$ GeV
   which is the left boundary point due to
   the internal  PYTHIA restriction on the
   lowest  value of the invariant mass of the
   initial state two-body system of any fundamental
   quark-parton  $2 \to 2$ subprocess.
   The spectrum finishes at 
   $M^{q \bar q}_{inv} = M^{l^{+}l^{-}}_{inv} \approx  2.5$ GeV.

\begin{figure}[!ht]
   \begin{center}
   \mbox{a)\includegraphics[width=7.2cm, height=5.2cm]{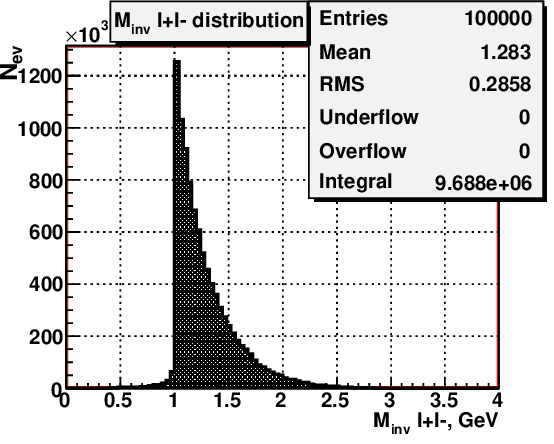}}  
   \mbox{b)\includegraphics[width=7.2cm, height=5.2cm]{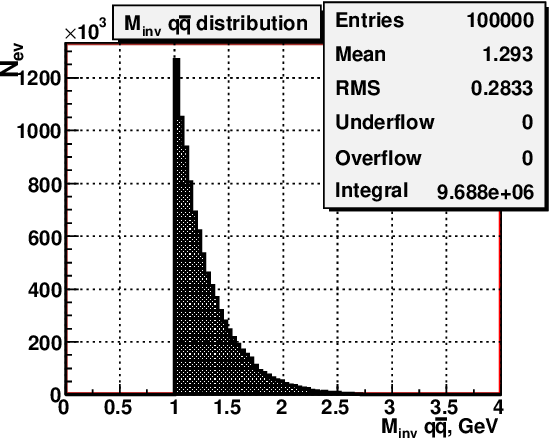}}  \\
     \caption{\small \it Distribution of number
                 of signal events 
		 versus the invariant  mass.
 {\bf a)} The invariant mass of  initial-state quark-antiquark
                       system $M_{inv}^{ q\bar{q} }$; 
{\bf b)}   The invariant mass of final-state lepton-antilepton
                       pair $M^{l^{+}l^{-}}_{inv}$.}		 
     \end{center}    
\vskip -0.5cm           
     \end{figure}  
		
   Different to the spectrum of the  invariant mass of the initial-state
   $q\bar{q}$-system,  the   invariant mass $ M_{inv}^{ l^{+}l^{-} }$ 
   of  the final-state  $l^{+}l^{-}$ system has a very small
   tail  at smaller than 1 GeV values $M^{l^{+}l^{-}}_{inv} < 1$ GeV 
   (see Fig.11{\bf {a}})
   \footnote{This left tail of $ M_{inv}^{ l^{+}l^{-} }$
   may appear  due to the final state radiation
   (FSR) of photons by the produced leptons.}.

    The  spectrum  of the total lepton pair    energy 
 ${E^{l^{+}l^{-}} = E^{l^{+}} + E^{l^{-}}}$ is  shown in  Fig.12{\bf {a}}.
    It is seen that  the lepton pair total
    energy distribution is by about 2 GeV longer 
    than the  spectrum of the fast leptons energy 
    $E^{l}_{fast}$ (see  Fig.3{\bf b}).

   \begin{figure}[!ht]
   \begin{center}
\vskip -0.5cm       
   \mbox{a)\includegraphics[width=7.2cm, height=5.2cm]{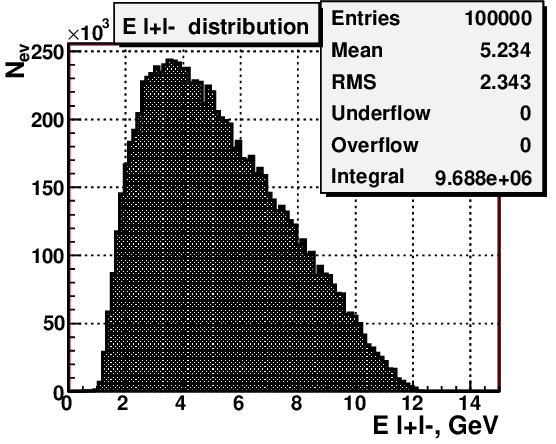}}  
   \mbox{b)\includegraphics[width=7.2cm, height=5.2cm]{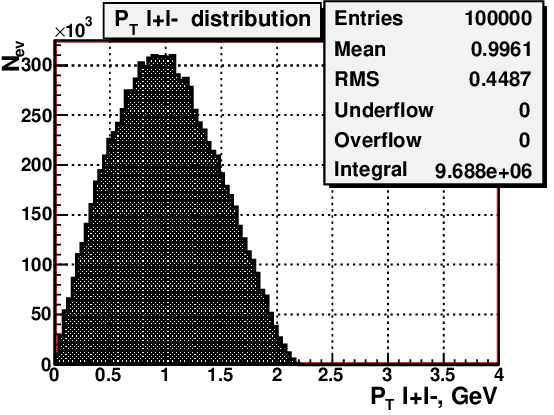}}  \\
     \caption{\small \it Distribution of number
                 of signal  events versus:
	       	 {\bf a)} the total lepton pair 
                  energy ${E^{l^{+}l^{-}}}$; 
	     {\bf b)} the total transverse momentum 
                     $P^{l^{+}l^{-}}_{T}$ of the 
                     lepton pair.}		 
     \end{center}    
\vskip -0.5cm           
     \end{figure}  
  
    The  mean value of the total energy of the lepton pair, as seen from 
    Fig.12${\bf {a}}$, is about 5 GeV.
From the same plot of  Fig.12 it is clearly seen that in more than a half 
 of events the lepton  pair energy lays in the interval 
      $4 \leq  {E^{l^{+}l^{-}} } \leq 12$ GeV. 
  So, one can conclude that the produced
 lepton pairs are rather energetic and 
     they can carry away quite a noticeable part
     of the total energy of  the colliding  $\bar{p}p$-system.

   The   distribution of the longitudinal  component 
 ${P^{l^{+}l^{-}}_{z} =   P^{l^{+}}_{z} + P^{l^{-}}_{z}}$
 of the total 4-momentum of lepton pair system (not shown here) has a shape
 which is  very similar to the energy ${E^{l^{+}l^{-}}}$ spectrum.       
 The explanation of this fact follows from the shape  of the  distribution of
 the modulus of  lepton pairs total transverse momentum 

\begin{equation}
  P^{l^{+}l^{-}}_{T} = |\vec{P}^{l^{+}l^{-}}_{T}| =
   |\vec{P}^{l^{+}}_{T} + \vec{P}^{l^{-}}_{T}|,
\end{equation}
which is presented as the 
``$P_{T}^ {l^+l^-}$ distribution'' in Fig.12{\bf {b}}.
    One may see that the distribution of the transverse component 
 $P^{l^{+}l^{-}}_{T}$ is  much more narrow  than that one of the 
    energy ${E^{l^{+}l^{-}}}$.  It covers
    the region $0< P^{l^{+}l^{-}}_{T} <2$ GeV, 
    like it was in a case of a single lepton distribution
    (see plots {\bf c} and {\bf d} of Fig.2). 
 $P^{l^{+}l^{-}}_{T}$  has a peak position 
at about 1 GeV, that is twice as large as 
    the analogous peak  position of a single
    lepton $P^{l}_{T}$  shown in the same plots {\bf c} and {\bf d} of Fig.2.
    Thus, we see that the main contribution to 
    the value of lepton pair energy comes from the
    longitudinal component    ${P^{l^{+}l^{-}}_{z}}$. 
 
 It is worth noting that according to Fig.1 the square of the invariant mass 
     $(M_{inv}^{l^{+}l^{-}})^{2} =Q^2 = q^2=
   ( P^{l^{+}} + P^{l^{-}})^2 $ has the meaning of the
   square of the momentum transferred from the 
   quark-antiquark pair to the lepton pair.
   Therefore, it plays the same role as the $Q^2$ in
   the processes of deep-inelastic scattering (DIS)
   of lepton over the proton. From 
   this point of view the diagram shown in Fig.1 looks 
   like the cross chanel analog of the diagram of 
   DIS  scattering of a lepton over the proton
   with the inclusive production of a proton in the
   final state. The essential difference is that in
   DIS case the momentum transferred is defined 
   by the relation $q_{dis} = P^{l}_{in} - P^{l}_{out}$
   ($P^{l}_{in}$ and $P^{l}_{out}$  are, 
   respectively, the  4-momenta of incoming
  and outgoing leptons)
   \footnote{ For DIS the value of $Q^2$ is defined
   as $Q^2 = -q_{dis}^2 = -(P^{l}_{in} - P^{l}_{out})^2$.},
   which differs from the definition of $Q^{2}$
   given above for the case of MMTDY process.
   Therefore, its square has the negative value 
   $q_{dis}^{2} \leq$ 0, while in MMTDY process  its square is positive: 
   $q^2 \equiv ( P^{l^{+}} + P^{l^{-}})^2  \geq 0$.
   
     To conclude this Section let us mention that the
   measurement of the invariant mass of lepton pair
   would allow to separate  the background to
    $J/\psi$  production events up to a good accuracy.

 \section{ Estimation of the size of $x-Q^{2}$ region 
         available for measurement  of   proton  structure function}

 ~~~   The distributions of Bjorken x-variables are
    shown in Fig.13  for up- (plot {\bf a}) and  down- (plot {\bf b}) quarks 
  \footnote{  The distributions  of antiquarks look similar to quark
    distributions for $\bar p p$ collisions}. 
 They represent the corresponding quark components of the proton structure
 function which was used in the present  simulation with CTEQ3L PDF.

   \begin{figure}[!ht]
   \begin{center}
   \mbox{a)\includegraphics[width=7.2cm, height=5.2cm]{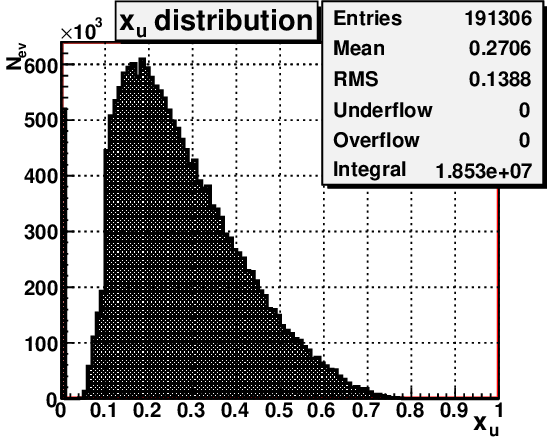}}  
   \mbox{b)\includegraphics[width=7.2cm, height=5.2cm]{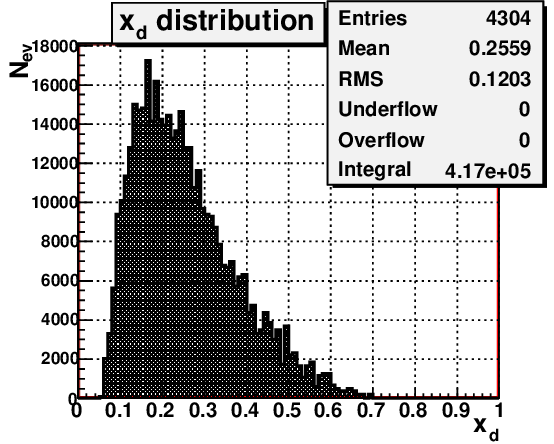}}  \\
     \caption{\small \it x distributions of valence:
	       	 {\bf a)} up quarks; 
	     {\bf b)}  down quarks.}		 
     \end{center}    
\vskip -0.5cm           
     \end{figure}  

  The most ineresting for us is the information about the size 
of the x-region which will be available at
 PANDA energies. We see from both
    plots of Fig.13 that x-varaiable spans the interval
    $0.05 < x < 0.7$. Recall that in  $\bar p p$ 
    collisions the transverse momentum $P_{T}$ plays 
    the role of the transferred momentum q.    
  
    So, combining the results from Figs.11 and 13 we can  conclude that, 
according to the  results of simulation with PYTHIA, we can hope to  get the 
 information about valence quark distributions in the kinematical region
 defined by the following boundaries:
    $0.05 \leq x \leq 0.7$ and $ Q^{2} \leq 6.2$ GeV 
     ($Q^{2} \equiv q^{2} = (P^{l^{+}l^{-}}_{T})^{2}$). 
    The important  point that  should be stressed here is that different
   to lepton-hadron scattering, which provides 
    the information about the structure 
    functions in the region of  negative, i.e. "space-like"
    values of the square of transversed  momentum
    $q_{dis}^2 = (P^{l}_{in} - P^{l}_{out})^2$, 
 the "annihilation" process $\bar{p}p \rightarrow l^{+}l^{-} + X$ allows
    to get the information about the structure
    functions in a region of  positive, i.e. "time-like" values
    of $q^2 \geq 0$.  Such a measurement of 
     quark distributions will be a good
    supplement to the planned measurement of
    proton elastic formfactor in the  region of "time-like"
    values of $q^2 \geq 0$ (see \cite{PANDATPR})
    and  the planned measurements of deep-inelastic
    process in a region of small values of $q^2 \leq 0$ at JLab and DESY.


\section{  Fake leptons in signal events}


 ~~~  The signal events, defined by the
    $q\bar{q} \rightarrow l^{+}l^{-}$ 
    subprocess, also contain  some hadrons in the 
    final state. Fortunately, their number is  essentialy
    restricted by the upper limit on the beam energy
    that may be available  at PANDA experiment.
    This circumstance may simplify greatly the 
    identification of  final state particles and the
    physical analysis due to  reduction  of the phase
    space and therefore to the reduction of the number of 
     hadrons  and other particles which may be   produced in event
    directly or in the  decays cascades of other hadrons. 
    These  hadrons may decay within the detector    volume
      \footnote{For the pion the $c{\tau}$ factor ($\tau$ is
      the mean life time of a particle, $\tau_{\pi}=2.6E-8$ sec) is equal to 
      $c{\tau}=7.8 $ meter according to PDG.}
    and thus produce the  background  leptons
    which may fake  the signal  leptons  ($\mu$, e) 
    produced in a  signal annihilation subprocess.
    In signal events with electron-positron pair
    producton  the fake electrons/positrons may appear also from muon 
    decays.  
    
    In this Section we shall consider 
    separatly the signal events with muon pair production
    (subsections 6.1 and 6.3) and the signal events 
    with electron pair production (subsection 6.4). 
    The subsection 6.2 includes the discussion
    of different kinematical distributions for charged
    pions which povide the main contribution to fake muons.
    Our analysis of fake leptons is based on 
    two  different  samples. One of them is the
    sample with signal muon pair events production
    while the second one with  electron pair
    production events.  This samples  were used in Section 2.

\subsection{  Fake muons}

~~~     We shall first discuss the case of background muons
    which may be produced additionally
    to a signal   "$\mu^{+}\mu^{-}$"- pair in the signal events 
     due to hadron decays.  For this reason, we shall call
    these fake muons also as ``decay  muons''
 (in the following we do not distinguish  $\mu ^{+}$ and $\mu^{-}$).
    The distribution  which shows the number  of the main  hadronic 
    ``parents'' of  muons in signal events 
    \footnote{ This kind  of information  can be extracted 
                    from  PYTHIA event listings.}
  is presented in the plot {\bf a} of Fig.14, while Fig.14 {\bf b} shows the 
distribution of muon ``grandparents'', i.e. the ``parents of muon parents''. 

    \begin{figure}[!ht]
   \begin{center}
   \mbox{a)\includegraphics[width=7.2cm, height=5.2cm]{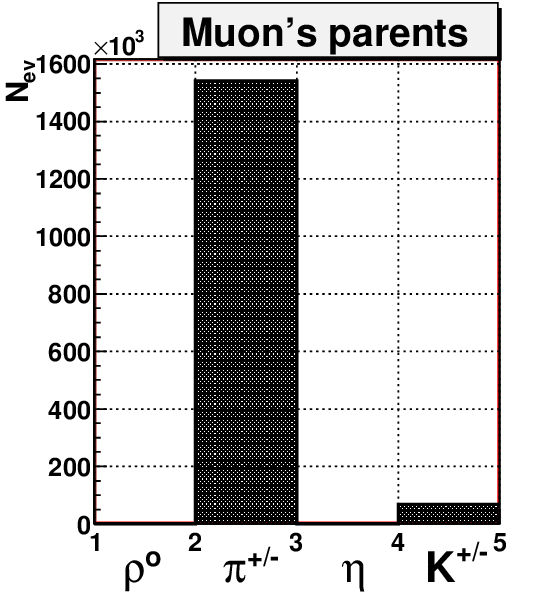}}  
   \mbox{b)\includegraphics[width=7.2cm, height=5.2cm]{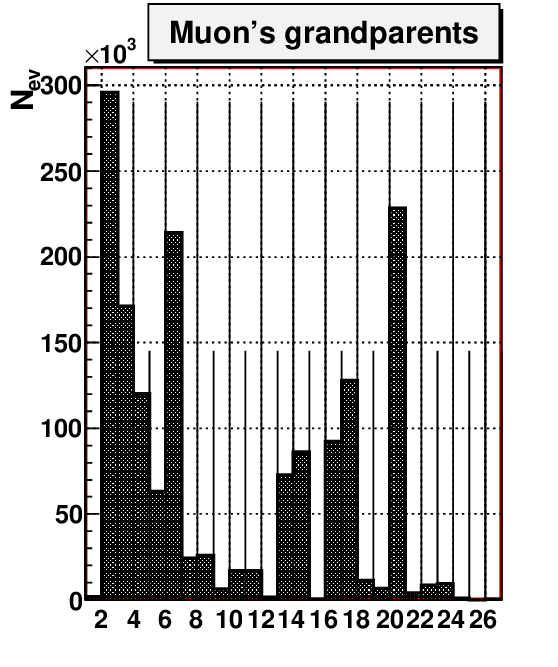}}  \\
     \caption{\small \it  Distributions of:
	       	 {\bf a)} parents of fake $ \mu$'s; 
	     {\bf b)} grandparents of fake $\mu$'s }		 
     \end{center}    
\vskip -0.5cm           
     \end{figure}     

    The correspondence between  the bin number
    on the  x-axis and  the name of the related grandparent
    of the muon can be found  from the right column of Table 7 of the 
     Section  11 (Appendix: Tables).
    \footnote{Sometimes, when the sea strange quarks take part
              in the fundamental  $q\bar{q} \rightarrow l^{+}l^{-}$
              interaction, a parent virtual K-meson appears 
              in PYTHIA  event listing together with the
              ``$K^{+/-}$ -like string'' (i.e. this string
              includes a strange quark) which  origin 
              flag points  onto the colliding proton  in the 
	      PYTHIA output listing. This 
              case do corresponds to the line which is named 
              as  ``$K^{+/-}$-like string '' in the 
               Table 7 of the Section  11 (Appendix: Tables). 
   It is seen from the plot Fig.14 {\bf a} that the  charged pions (bin 2)
   deliver the main decay muons  background while the contribution of 
   $K^{\pm}$-mesons (bin 4}) is of about  one order less.
   From Fig.14 {\bf b} and the right column of Table 7 
   of the Section 11 (Appendix: Tables)  one can conclude that the strings 
   (bin 2 in Fig.14 {\bf b}), $ \omega $ (bin 6) and
   $\Lambda^{0}$ (bin 20) are the main grandparents of muons.
   It is of interest to consider in more detail the
    kinematical  distributions of  charged pions 
    as the main source of fake muons and to 
    compare them with the distributions of produced
    muons. We shall do it in the following subsection.
  

\subsection{Kinematical distributions of parent pions}

         
~~~     The  left-hand side of Figure 15 includes (like Figs.2  and 3) 
three plots ({\bf a, c, e}) containing the distributions (top to bottom)
    of the  number of  events versus the  energy  $E_{\pi}$ 
   (plot {\bf a}),  the transverse  momentum  ${PT_{\pi}}$ 
    (plot {\bf c}) and  the  polar angle  $\theta_{\pi}$ 
    (plot {\bf e}) of charged pions which appear
in the sample of generated  by PYTHIA signal 
muon events (100 000) based on the 
 quark level subprocess    $q+\bar{q} \to \mu^{+}\mu^{-}$ (this
 sample was discussed in the Section 2 and in the begining of this Section). 

  \begin{figure}[!ht]
     \begin{center}
\vskip -0.5cm       
   \mbox{a)\includegraphics[width=7.2cm, height=5.2cm]{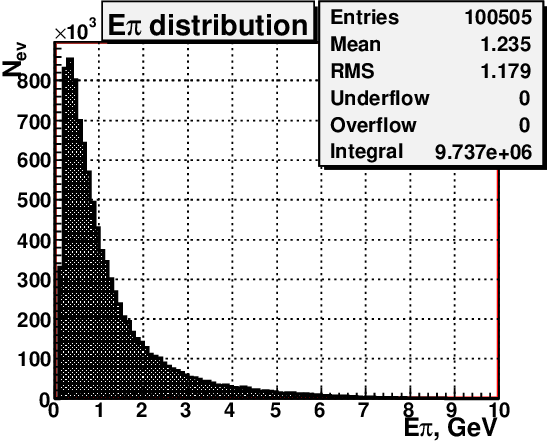}}  
   \mbox{b)\includegraphics[width=7.2cm, height=5.2cm]{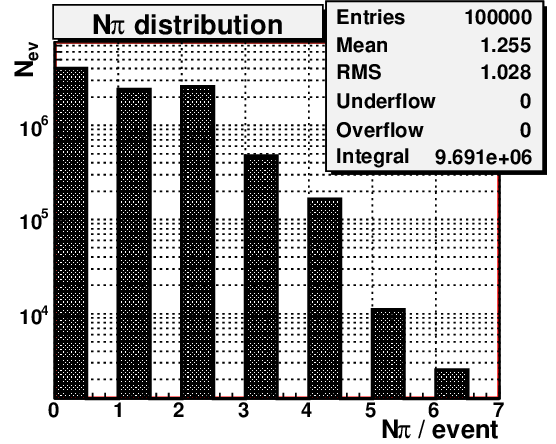}}  \\
   \mbox{c)\includegraphics[width=7.2cm, height=5.2cm]{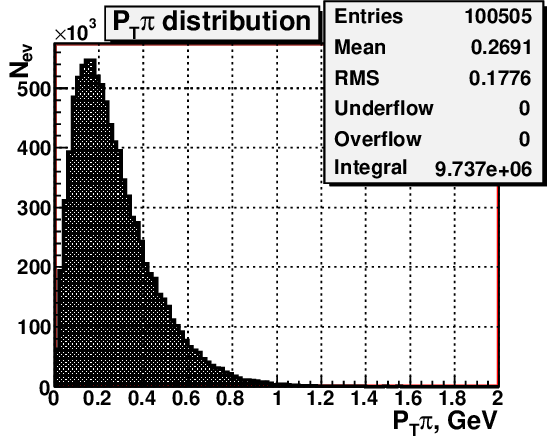}}  
   \mbox{d)\includegraphics[width=7.2cm, height=5.2cm]{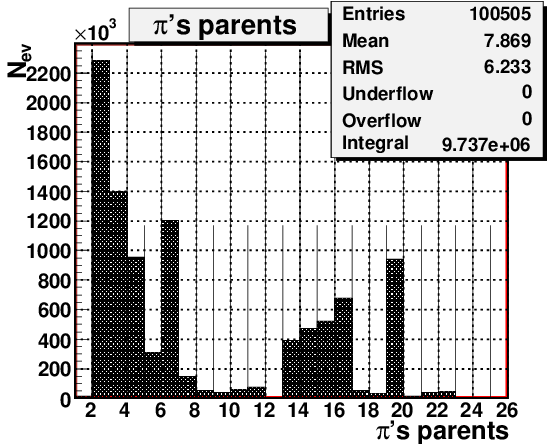}}  \\
   \mbox{e)\includegraphics[width=7.2cm, height=5.2cm]{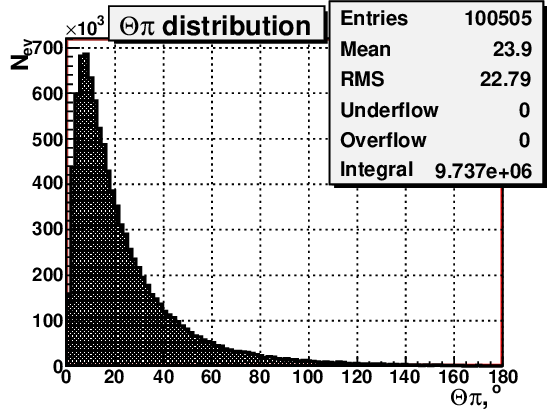}} 
   \mbox{f)\includegraphics[width=7.2cm, height=5.2cm]{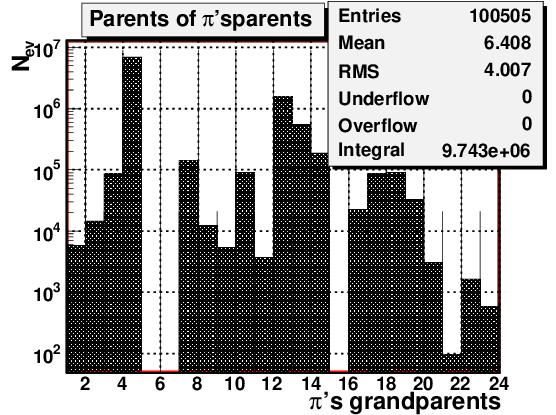}}  \\
		      		          
     \caption{\small \it Left column:   
              distributions of the number of
              pions versus {\bf a)}: the energy $E_{\pi}$; 
             {\bf c)}: transverse momentum  $P_{T}{\pi}$  and 
             {\bf e)}: polar angle ${\theta_{\pi}}$
              of the produced pions.
	       Right column: 
             distributions {\bf b)}: of the total number
             $N_{\pi}$ of charged ${\pi}$-mesons in the signal events; 
	     {\bf d)}: of the particles that give birth
	        to the pions ($\pi$'s parents) and {\bf f)}: of the 
		particles which are parents to $\pi$'s parents}	 
     \end{center}    
\vskip -0.5cm           
     \end{figure}

    The plot {\bf b} in the same Fig.15 shows the distribution
    of  number $N_{\pi}$ of charged ${\pi}$-mesons
    produced per event in generated signal events. 
    The  first left   bin in this  plot shows the number of 
    events  without charged pions. One may see 
    that there is a huge number of signal events 
    (about $42\%$) which \textit {do not contain  at all any 
    charged pions} (the final states in  the most part of
    these events, as it  can be  seen from the PYTHIA
    event listings, do include mostly  nucleon-antinucleon 
    pairs).  Thus,  with a good accuracy,  we may expect  
     that {\it about  $42\%$ (taking in account charged kaons 
    decays) of events,  which  include the signal muon
    pairs,  will not  contain any additional  fake decay  
    muons}.  Let us underline that  PYTHIA provides 
    a good (at least one of the best if not 
    the best) but still a model approximation 
    \footnote{Due to the fact that there is no complete
              physical and theoretical understanding of 
              parton-to-hadron fragmentation processes, so far.}
   to the hadronization processes.

    From the second and the third bins of $N_{\pi}$ 
    of the same  plot {\bf b}  one sees  that  about  
    $24\%$ of signal events may  have only one charged  pion  
    and about  $27\%$ of events may have  two charged pions
    in the final states.   The  other bins of  the same plot 
     {\bf b}  demonstrate that about
    $5\%$ of events may  have three charged pions, about $1.5\%$ 
    of events do contain 4 final-state   charged  pions and there is
    a very small fraction of events  containing from 5 to
     6 final-state pions
    \footnote{there is only one event  which includes 7 pions
                    is found among 100 000 of generated events.}.
        
    The origin of all produced pions 
    may be seen from the  plot {\bf d}  (``$\pi$'s parents'')
    of Fig.15, where each bin on the x-axis  corresponds to the 
    parent particle of a pion 
    (one can find the  correspondence  of the number of the bin 
    to  the name of a parent particle using  the left-hand column
    of  Table 8 of Section  11 (Appendix: Tables). From this plot 
    {\bf d}  and Table 8 one can see that the dominant source 
    (bin 2) for pion production  are the  strings
     (about $22\%$)  which are one of the main 
    objects in LUND  fragmentation model. The bins 3-6 do 
    correspond,  respectively, to the  $\rho$-, $\eta$- 
    and $\omega$-mesons. The decays of these light 
    vector mesons give quite a noticeable  contribution 
    (about $39\%$ of all  entries). The  contribution
    of parent K-mesons (bins 7-10)  is close to 
    $4\%$ of all entries. The next  sizable portion of pions
    (about $21\%$ of entries)  comes  from the family of 
    $\Delta$-resonances (bins 13-16) and from decays of
    $\Lambda^{o}$  (see the corresponding 19-th bin which
    contains   about $10\%$ of entries).

    In the same way, the corresponding  distributions of pion's grandparents
    are shown in the plot {\bf f}  of Fig.15 (see also 
    the  right-hand column of  Table 8 of  Section 11,
    (Appendix: Tables). One can see   that among the grandparents
    the denominating position  belongs to the strings (their
    4-th bin contains more than  $74\%$ of entries).
    Then  follow the bins which include  K-mesons 
    (bins 7-9) and also  $\eta^{'}$- and $\phi$- mesons  (bins 10 and 11)
    which all together  include about    $2.5\%$  of
    events. A new object in this grandparents plot, 
    comparing to that one of the parents, is  a 
    group of diquarks (bins 12-14), which total contribution
    is  about $18.5\%$ of entries. The group of $\Sigma$- and
$\Xi$-resonances (bins 16-23) gives a bit more than $3.5\%$ of pions parents.


\subsection{Kinematical and vertex distributions of
                     fake decay muons in signal events }

 
~~~~    Fig.16  includes a set of plots with the
    distributions of background fake decay muons, which 
    are  contained in the generated sample of 100 000
    signal muon events  described in the Section 2.
    This decay muons come from 
    all possible decay channels of produced hadrons
    including the pion decays which were discussed above.
    It is clear that not all hadrons will decay within
    the detector volume. Therefore, in this subsection 
    we have used the existing PYTHIA option which allows
    to take into account the restricted decay volume.
    As for the first approximation for a real PANDA 
    detector volume we have chosen the cylinder with 
    R=2.5m and L=8m. Because of this the number of 
    entries in all the plots of Fig.16 (exept the plot 
    {\bf b}, which shows the number  of  all generated 
    events) is equal to 16601.  It means 
    that the fraction of signal processes, which 
    include fake muons, reduces to about 16.6 $\%$ 
    after taking into account the detector size.

  \begin{figure}[!ht]
     \begin{center}
\vskip -0.5cm       
   \mbox{a)\includegraphics[width=7.2cm, height=5.2cm]{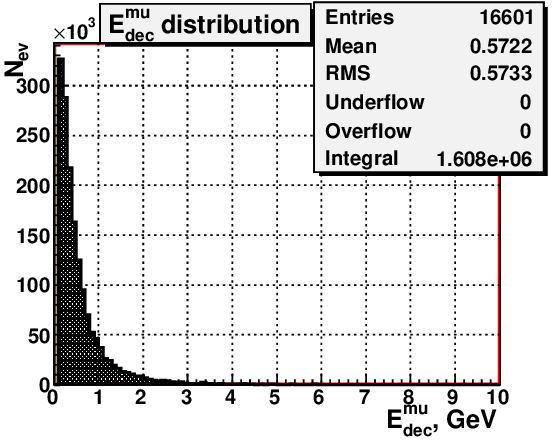}} 
   \mbox{b)\includegraphics[width=7.2cm, height=5.2cm]{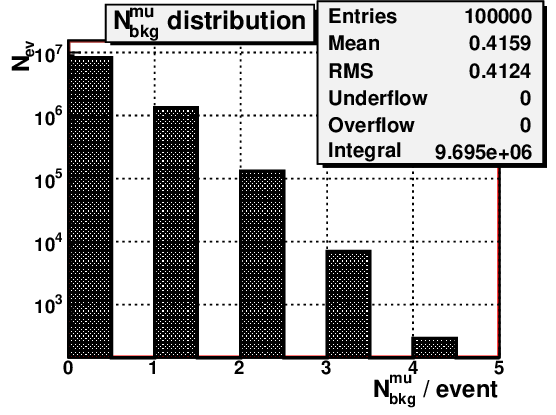}}  \\
   \mbox{c)\includegraphics[width=7.2cm, height=5.2cm]{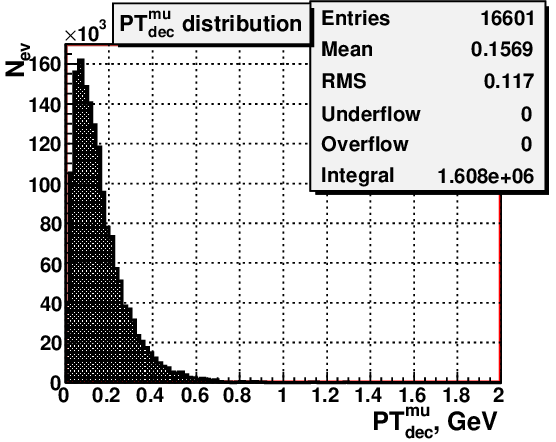}}  
   \mbox{d)\includegraphics[width=7.2cm,
 height=5.2cm]{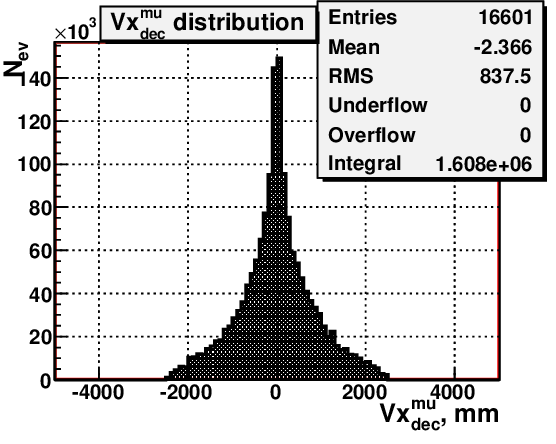}}  \\
   \mbox{e)\includegraphics[width=7.2cm, height=5.2cm]{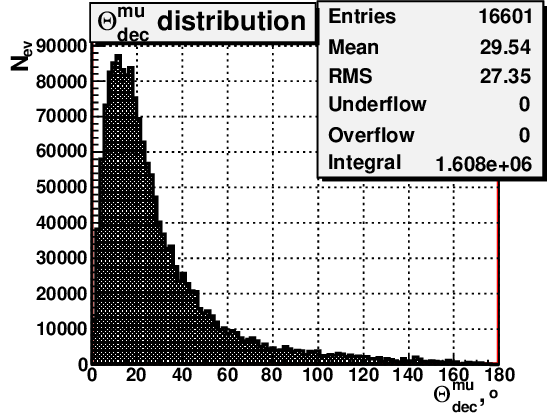}}  
   \mbox{f)\includegraphics[width=7.2cm, height=5.2cm]{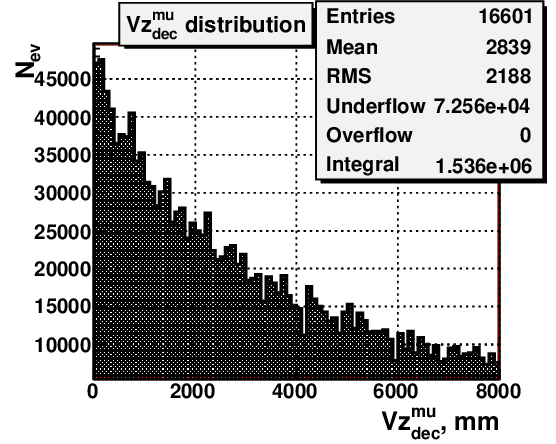}}  \\
		      		          
     \caption{\small \it Left column:   
              distributions of the number of
              decay muons versus their {\bf a)}: the energy $E^{\mu}_{dec}$; 
             {\bf c)}:  transverse momentum  $PT^{\mu}_{dec}$  and 
             {\bf e)}:   polar angle $\theta^{\mu}_{dec}$.            
	       Right column: 
    distributions of {\bf b)}: the total number $N^{\mu}_{bkg}$  of fake
             muons per the signal event; 
	    {\bf d)}: x coordinate of the fake muon production vertex
           and {\bf f)}: z coordinate of the fake muon production vertex }
	 
     \end{center}    
\vskip -0.5cm           
     \end{figure}

    The distribution of a number of muon signal events versus
    a  number of fake decay muons $N^{\mu}_{bkg}$, contained
    in each event,  is shown in the  plot {\bf b}. It is seen
    that there can be up to 4 muons in
    the final state.  From the first bin of this
    plot  one may see that
    \textit { about $83\%$ of events have no background fake  muons at all.}
    This value agrees well with the number   of entries shown in 
    the plot {\bf a} which was discussed above.
    The amount of events free of fake muons 
    does not corresponds to the  previously
     discussed amount of the events without charged
        pions  because of applied restriction of the  decay volume.
     It is also seen from the plot {\bf b} 
    that the numbers of events with one
    fake muon, with two and three fake muons are
    about by, respectively, one, two and three 
    orders less than the number of events without
    any fake muon.

     The left column of Fig.16 includes (from top to bottom) 
    the energy  $E^{\mu}_{dec }$, transverse momentum
    $PT^{\mu}_{dec }$  and  polar angle $\theta^{\mu}_{dec }$
    distributions of  muons which appear from hadron decays. 
    From comparison of these plots with their 
    analogs from Fig.2, i.e. of those which are
    done for the signal muons, one may see that 
    the fake muons are less energetic than the 
    signal ones. One may also find  that the mean
    value of the  signal muons energy (see plots {\bf a, b}  of Fig.2)
    $<E^{\mu}> $ = 2.6 GeV  corresponds to  such a
    point in the energy spectrum of fake muons 
     which decay in the restricted volume,
     where the contribution of fake muons 
    (in the same signal events) is very low. 
    Analogously, the mean value of  the PT-distribution
    of signal muons  $<PT^{\mu}_{signal}> = 0.7$ GeV 
    (see Fig.2)   corresponds to that point where the
    spectrum of  $PT^{\mu}_{dec}$ practically  vanishes.
       
      Therefore,  one has to look for the set of
    some reasonable cuts (chosen with an account
    of the real detector effects) on a muon energy  $E^{mu}$ 
    as well as on its  $PT^{mu}$ value, which may 
    lead to an essential reduction of decay muons  
    contribution and to keep at the same time the
     main part of signal events. For example,
     the comparison of the  plots {\bf a} and 
    {\bf c} of the Fig.2 with the  analogous plots
    in Fig.16,  leads to a conclusion that the   cuts
     $E^{\mu} > 0.2$ GeV,   $PT^{\mu} > 0.2$ GeV 
    may allow  to get rid of about  $66\%$ of 
    decay muons in the signal events and  to save the most
    of signal (i.e. belonging to  signal
    $\mu^{+}\mu^{-}$- pair) slow and fast muons.
    We shall return to this problem a bit later.
     
      Another way  which may help to discriminate 
    the signal muons from the "decay" ones  is
    to use  the information about the position of
    fake muon production vertex  and the reconstruction
    of the invariant mass of the parent and grandparent
    hadrons. The plots  {\bf d} and {\bf f} 
    of Fig.16 contain  the distributions 
    of Vx- and Vz- (z-axis is chosen  along the beam 
    direction)  components of the  3-vector 
    $V=(Vx, Vy, Vz)$, which gives the position  of a 
    fake muon production vertex in millimeters (mm)
  Let us note that these plots are obtained within the  PYTHIA 
level of simulation, i.e. without  an account of  details of detector
 construction and the effects caused by the  magnetic field. 
 Withing this approximation the  distributions of  Vx- and
    Vy- components have to  be similar.  For this reason, 
    we   shall show in the following only the distribution
    of Vx- component.
    Seen   from the  Fig.16 the Vz 
    component of about $3-5\%$ of events may be 
    rather close to zero, i.e. to the interaction
    point. The corresponding fake muons, produced 
    near the interaction point, may give rise to 
    the most difficult background.
    The  contribution of the background muons from 
    the other type of events (based mainly on minimum 
    bias and QCD partonic subprocesses)  will be  
    discussed in the following  Section  7.


\subsection{  Fake electrons }


~~~    Now let us  consider the situation with 
   fake decay electrons and positrons
   background in the  case of  signal  processes 
  based on the quark level subprocess of electron-positron pair production 
   \footnote{ In the following we shall  use ``electron'' as
              a common name for the background electrons and
              positrons.}.
    Fake electrons produced in these decays will be also
   called in the following  as  ``decay'' ones.
    Recall, as it was already mentioned in the 
    Section 2 and in the begining of the present one, that
     we  shall use the sample of 100000 
    generated signal events with $e^{+}e^{-}$ production.

   Fig.17{\bf a} presents the contribution of different
    parent particles  into  the process  of creation of the fake electrons.
   It clearly demonstrates that among all  shown
   sources of fake electrons the contribution  of neutral
   pion  $\pi^{0}$ decay  (bin 2) is a dominat one. It provides 
   a much higher (about of one order) contribution than all of
   other decay channels.
     The electrons/positrons may appear in neutral pion
   decay only through the channel of Dalitz decay 
   into a photon and  an electron-positron  pair:
   $\pi^{0} \rightarrow \gamma + e^{+}e^{-}$.  
    The next contribution, which is by about one 
   order less, give, in decending order, muon  (bin 1), $\eta$-
   and K-mesons decays as well as the decay of   $\Lambda^{0}$.
   Neutral pions in their turn may  arise from decays
   of $\eta$- and  $\omega$-mesons or heavier mesons
   and  baryons, produced as  resonance
   states according  to the  LUND fragmentation model.

   \begin{figure}[!ht]
   \begin{center}
   \mbox{a)\includegraphics[width=7.2cm, height=5.2cm]{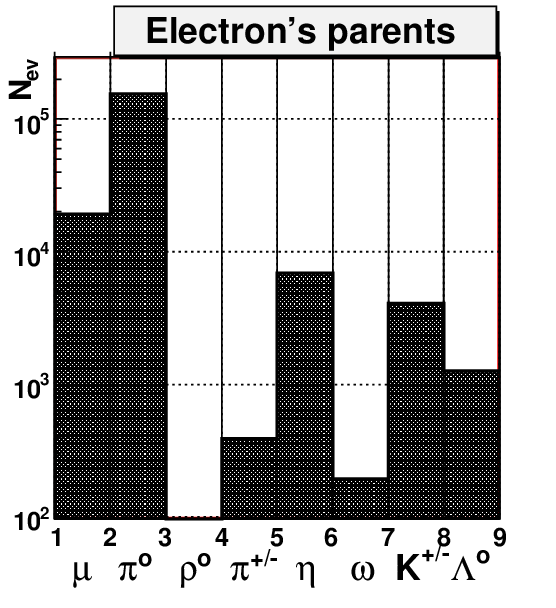}}  
   \mbox{b)\includegraphics[width=7.2cm, height=5.2cm]{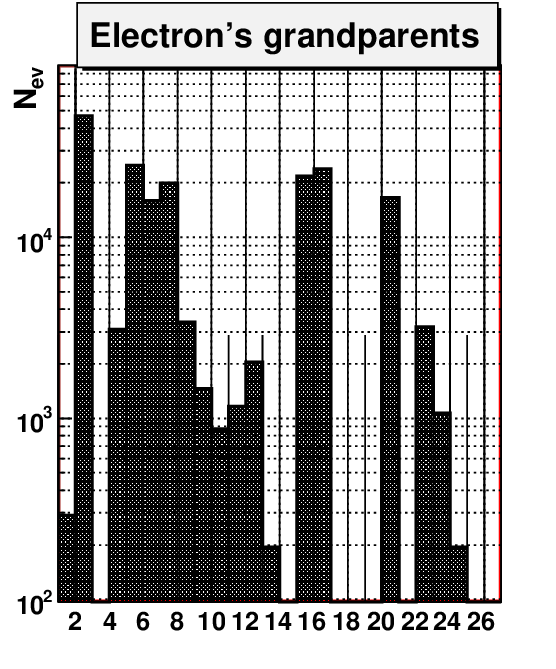}}  \\
     \caption{\small \it  Distributions of:
	       	 {\bf a)} parents of fake $e^{+}/e^{-}$; 
	     {\bf b)} grandparents of fake $e^{+}/e^{-}$ }		 
     \end{center}    
\vskip -0.5cm           
     \end{figure}

   Fig.17 {\bf b} shows the  distribution of a number   of 
   background electrons versus the type of their 
   grandparents. The correspondence  between the 
   bin numbers shown on the x-axis  and the
   name of the grandparent particle can be  found
   from the  left column of the Table  7 of 
   Appendix: (Tables which are presented in the Section  11).
   From comparison with this Table one can see that the
   strings (bin 2) are the main source of electron parents. 
   Then follow a group of $\rho$-, $\eta$- and 
   $\omega$-mesons (bins 5-7) as well as the group of
   $\Delta$-resonances (bins 15-16) and $\Lambda^{o}$ (bin 20).
   Two orders less contribution is provided by the group of
   kaons (bins 8-11) which is followed by $\eta^{'}$ 
   resonance (bin 12)
   and the group of $\Sigma$-resonances (bins 22-24).

    The plots, containing the distributions  of the 
   fake  decay electrons which appear in 
   signal events (based on the  $q\bar{q} \to  e^{+}e^{-}$ subprocess) are  
   presented  in Fig.18.
    They are done basing on the sample of 100 000 
   generated by PYTHIA signal events  which were
   discussed in the Section 2 and contain  signal $ e^{+}e^{-}$ pairs.
     Like in previouse subsection we use here the
   same restriction on the detector volume.

  \begin{figure}[!ht]
     \begin{center}
  \vskip -0.5cm       
   \mbox{a)\includegraphics[width=7.2cm, height=5.2cm]{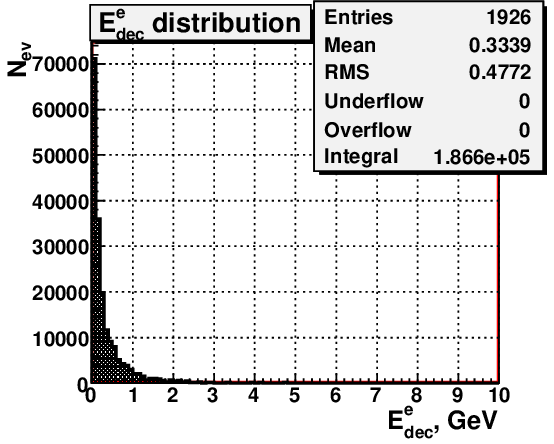}}  
   \mbox{b)\includegraphics[width=7.2cm, height=5.2cm]{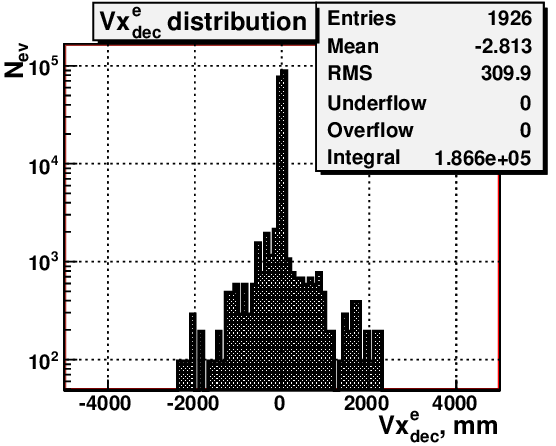}}  \\
   \mbox{c)\includegraphics[width=7.2cm, height=5.2cm]{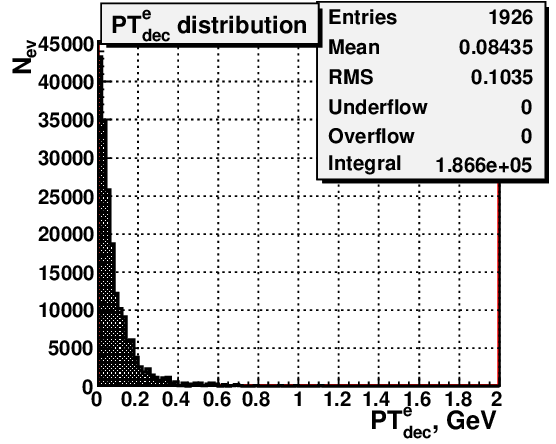}}  
   \mbox{d)\includegraphics[width=7.2cm, height=5.2cm]{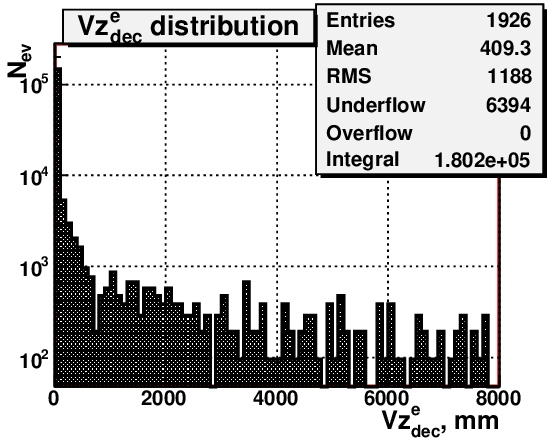}}  \\
   \mbox{e)\includegraphics[width=7.2cm, height=5.2cm]{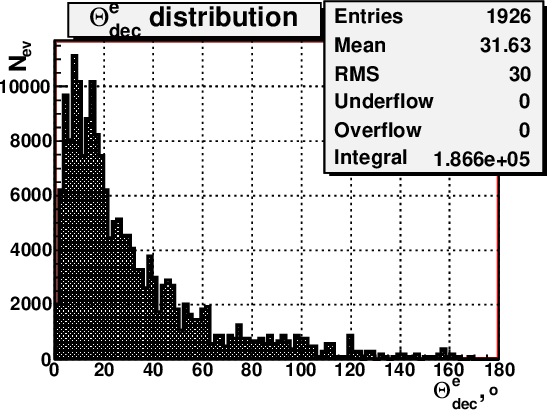}}  
   \mbox{f)\includegraphics[width=7.2cm, height=5.2cm]{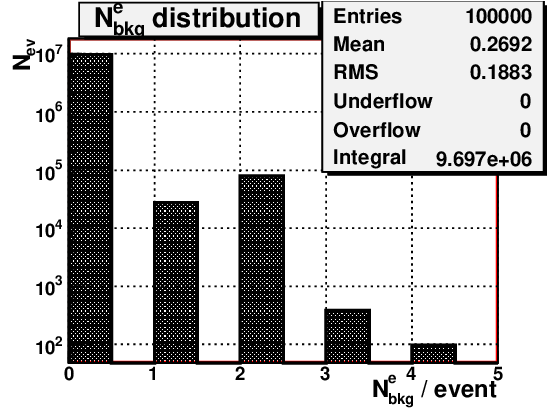}}  \\
     \caption{\small \it Left column:   
              distributions of the number of
              decay electrons versus their:
  {\bf a)}:the energy $E^{e}_{dec}$; 
             {\bf c)}:  transverse momentum  $PT^{e}_{dec}$  and 
             {\bf e)}:   polar angle $\theta^{e}_{dec}$.            
	       Right column: 
    distributions of {\bf b)}: the total number $N^{e}_{bkg}$  of fake
             muons per the signal event; 
{\bf d)}: x coordinate of the fake $e^{+}/e^{-}$ production vertex
   and {\bf f)}: z coordinate of the fake $e^{+}/e^{-}$ production vertex }
 \end{center}    
\vskip -0.5cm           
     \end{figure}

    From the  statistics frame in the upper part of
   plots one may see that in a case of decay electrons 
the  number of entries is 1926 
(\textit i.e. the fraction of the  signal processes 
   which includes fake electrons   is about $2\%$).

    The comparison of the energy $E^{e}_{dec}$ and transverse
    momentum  $PT^{e}_{dec}$ distributions of fake electrons
    in signal events, which are given in the left-hand column 
    of Fig.18, with the analogous plots for fake muons from the
    left-hand column  of Fig.16, allows to conclude that in
    electron case the distributions fall at least twice steeply
    than in muon one.

    Analogously with the previous subsection, the 
    comparison of the same  $E^{e}_{dec}$ 
    and transverse $PT^{e}_{dec}$ distributions of fake electrons
    in signal events, shown in plots {\bf a} and {\bf c} of Fig.18,
    with the plots {\bf a} and {\bf c} of Fig.2 for signal
    electrons leads to the conclusion that the soft cuts  like  
    $E^{e} > 0.2$  GeV  and  $PT^{e} > 0.2$ GeV  may
    allow to  eliminate the most of  fake electrons 
    at the cost of about $10\%$ loss of signal events.

   The right-hand  column of Fig.18 includes plots {\bf b} and {\bf d}. They 
   contain the values of the  Vx- and Vz-
   components of the 3-vector V  which points
   the  position of electron production  vertex.
    In contrast to the form of vertex distribution for 
muon production case (see the plot {\bf d} in Fig.16),
   the concentration of Vz-component of electron production 
   (see plot {\bf d} in Fig.18)
   near the interaction point  poses the essential
   difference of a $e^{+}e^{-}$ channel as compared to 
   a case of  $\mu^{+}\mu^{-}$  channel. In the last case,
   the most of background muons are produced by light
   charged pions which may  decay in flight at a 
   rather  large distance from the interaction point.
   It is also seen from plot {\bf f} in Fig.18 and the 
   analogous plot {\bf a} in Fig.16 that  the number of fake 
   electrons or fake muons in signal events 
   may be up to four in both cases.

\subsection{ Cuts for fake leptons reduction in signal events}     

~~~ The analysis of distributions discussed above
    leads to the conclusion that the following cuts:
  
$\bullet$ 
  1) we select the events with the only  two leptons 
        with  $E_{l} \geq$ 0.2 GeV,  $PT_{l} \geq$ 0.2 GeV; 
	
$\bullet$
 2)   the charges of these two leptons must   be of the opposite sign; 

$\bullet$  
 3)   the vertex of lepton origin lies  within the range 
       $R_{vtx} \leq$ 15 mm from  the interaction  point; \\

   which, being  applied to the  sample of signal events, can
  allow one to select a subsample  which  include 
  a strongly reduced fraction  (fr)  of events  containing 
  fake leptons ($fr_{e}=0.008\%$ for the case of electron pair
  production  and $fr_{\mu}=0.001\%$ for the muon pair
  production). The loss of the signal events due to
  application of  cuts  1)--3) is shown in Table 3.
         
\begin{table}[!hbt]
\vskip -0.5cm  
\caption{ Loss of the signal events (in $\%$) after 
              application of cuts 1)--3)}
\begin{center}
\begin{tabular}{|c|c|c|}              \hline
 N of cuts & $e^{+}e^{-} production$ & $\mu^{+}\mu^{-} production$ \\\hline
    1       &    $14.330\%$       &    $16.525\%$      \\\hline  
    1 $\&$ 2       &     $14.340\%$        &  $16.805\%$       \\\hline  
    1 $\&$ 2 $\&$ 3       &    $14.341\%$        &   $17.108\%$  \\\hline  
\end{tabular}
\end{center}
\end{table}

 One can see from this Table that it is  possible to select
 the signal events  which are almost  free of  background fake  leptons
 at the cost of  diminishing of the signal events sample by
 $\approx$ 17 $\%$ for $\mu^{+}\mu^{-}$ and 
 $\approx$ 14$\%$ for $e^{+}e^{-}$ production.


\section{  QCD and minimum-bias background events}


 ~~~  The other source of the background is the leptons 
   produced in the minimum-bias (low - $P_{T}$  and
    diffractive scattering) events and 
   QCD  background  these are mainly  
   $q + g \rightarrow q + g$,  $g + g \rightarrow g + g$ 
   and $q + q' \rightarrow q + q'$) processes, where
   the possibility of appearance of two 
   (and more) leptons in  the final state  is very high. 
    For analysis of these processes $10^6$ events of antiproton
    diffraction over proton target with $E_{beam} = $14 GeV were generated
    with PYTHIA 6.4. These events include mentioned above processes,
   including the signal one  $\bar q +  q \to l^{+}l^{-}$. 
    According to PYTHIA the total cross section of these  processes
    $\sigma^{bkg}_{tot} = 50.17$mb   is about $10^7$
    times higher than the cross section of the signal
     MMTDY subprocess    $q + \bar{q} \rightarrow l^{+} + l^{-}$: 
     $\sigma^{\bar q q \to l^+ l^-}  = 5.57 \cdot 10^{-6}$mb.  
 In the following subsections 7.1-7.2 we shall present
    the distributions obtained without use of any cuts.
           
\subsection{  Muon background.}

~~~~   The distribution  of the parents of 
    muons  produced in background   minimum-bias and QCD events 
 is presented  in Fig.19 {\bf a}.  It is seen that, like
   in a case of fake muons in signal events  (see
   Fig. 14 {\bf a}), the main contribution comes
   from $\pi^{\pm}$- and $K^{\pm}$-meson decays. 
    Fig.19 {\bf b}
   \footnote{ which is slightly different from its 
              analog in Fig.14 for the signal events.}
   shows the distribution of muons grandparents.
   One can see (using  the right-hand column 
   of Table 7 of the Section  11 (Appendix: Tables)
   that the main grandparents of muons in QCD and 
   minimum-bias events are the clusters and strings
   (bins 1, 2) as well as the $\rho$-, $\eta$- and 
   $\omega$-mesons (bins 3-6). Then follow the
   $\Delta$-resonances (bins 16, 17) and $\Lambda^{0}$ (bin 20).

   \begin{figure}[!ht]
   \begin{center}
   \mbox{a)\includegraphics[width=7.2cm, height=5.2cm]{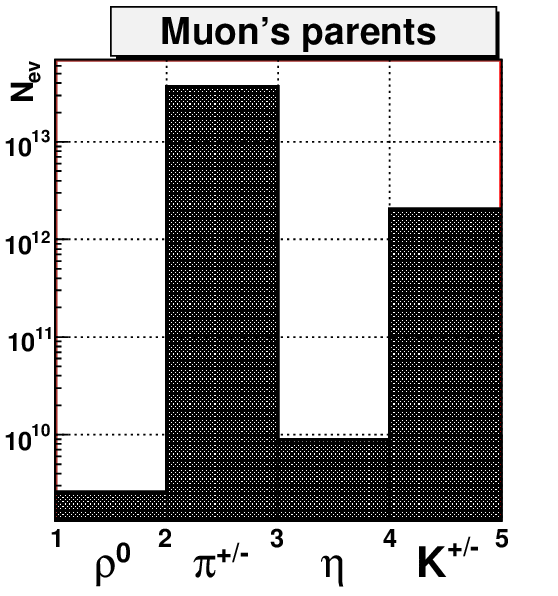}}  
   \mbox{b)\includegraphics[width=7.2cm, height=5.2cm]{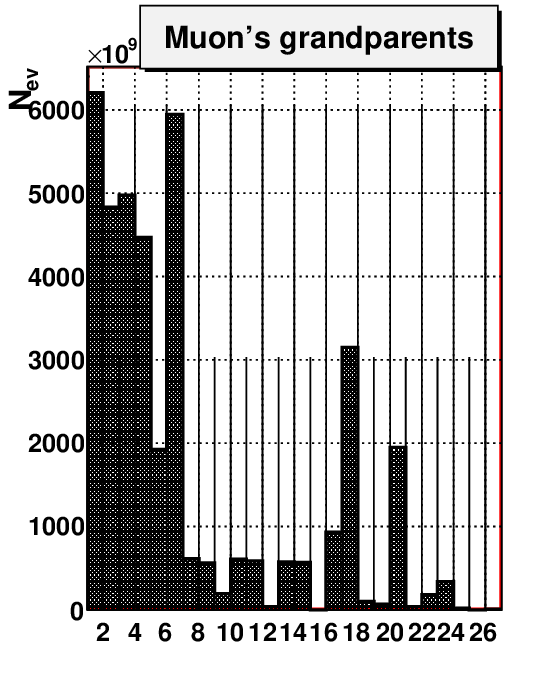}}  \\
     \caption{\small \it  Distributions of:
	       	 {\bf a)} parents of background $\mu^{+}/\mu^{-}$; 
	     {\bf b)} grandparents of background  $\mu^{+}/\mu^{-}$ }	 
     \end{center}    
\vskip -0.5cm           
     \end{figure}  
   
    The  kinematical and other distributions of 
   muons produced in the above mentioned generated
  background minimum-bias and QCD events  are shown in Fig.20 
   It is seen that the kinematical distributions
    (plots {\bf b, c, e}) do not differ so much (that is natural) from 
   those of fake  ``decay'' muons produced  in the signal 
   $p\bar{p} \rightarrow l^{+}l^{-} + X$  processes (see Fig.16). 
  
  \begin{figure}[!ht]
     \begin{center}
\vskip -0.5cm       
   \mbox{a)\includegraphics[width=7.2cm, height=5.2cm]{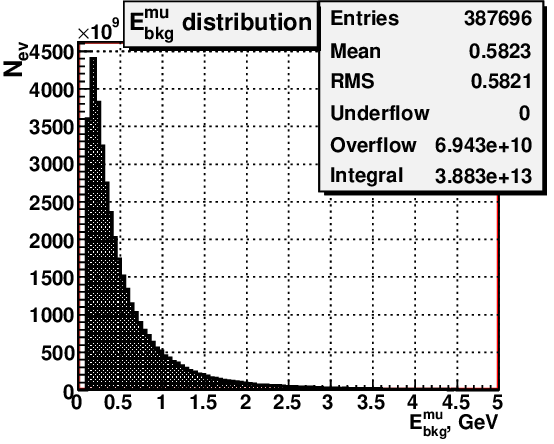}}  
   \mbox{b)\includegraphics[width=7.2cm, height=5.2cm]{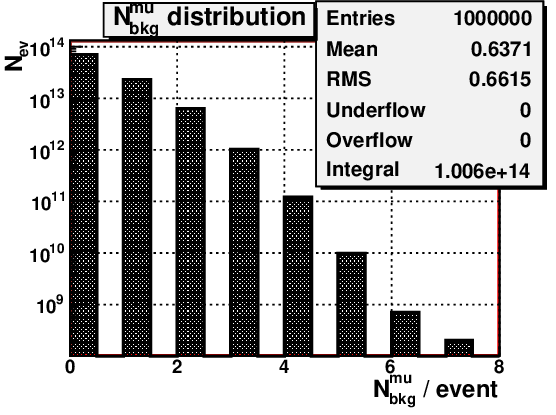}}  \\
   \mbox{c)\includegraphics[width=7.2cm, height=5.2cm]{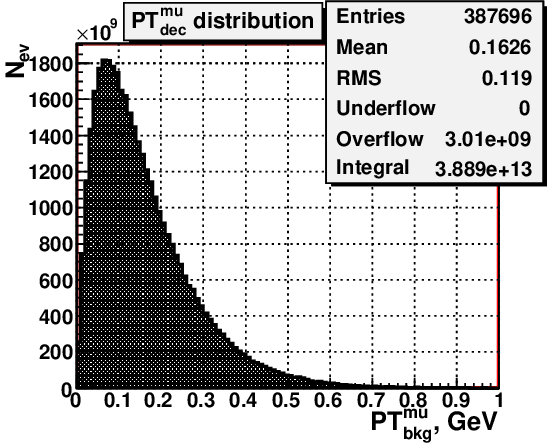}}  
   \mbox{d)\includegraphics[width=7.2cm, height=5.2cm]{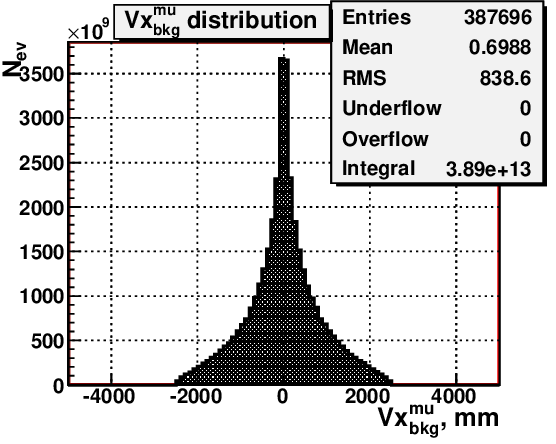}}  \\
   \mbox{e)\includegraphics[width=7.2cm, height=5.2cm]{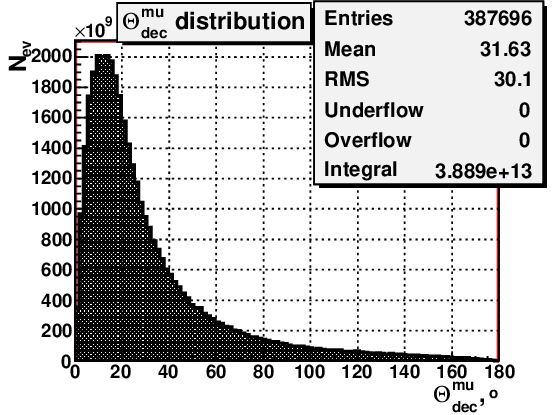}}  
   \mbox{f)\includegraphics[width=7.2cm, height=5.2cm]{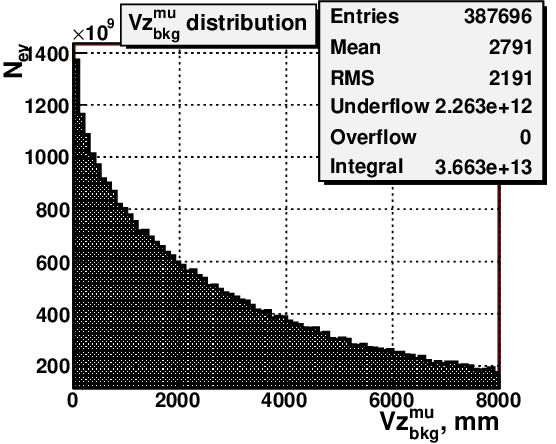}}  \\		    
     \caption{\small \it Left column:   
              distributions of the number of
              background muons versus 
 {\bf a)}:    energy $E^{\mu}_{bkg}$; 
{\bf c)}:    transverse momentum  $PT^{\mu}_{bkg}$  and 
              {\bf e)}:   polar angle $\theta^{\mu}_{bkg}$.            
	       Right column:   distributions 
 {\bf b)}:    of the total amount $N^{\mu}_{bkg}$  of   muons per  event; 
 {\bf d)}: x coordinate of the background 
               muon production vertex and 
{\bf f)}: z   coordinate of the background muon production vertex }
	      \end{center}    
\vskip -0.5cm           
     \end{figure} 

   The distribution of the number of generated 
   background events versus the amount of fake 
   muons produced  per event, i.e. $N^{\mu}_{bkg}$,
   is shown in plot {\bf b)} of Fig.20. 
   It  differs noticeably from its analog shown in Fig.16
   which  contains only  the distribution of fake "decay"
   muons in signal lepton pair production events. 
   The number of muons in the final state,  contained in
   one and the same background 
   event, can be up to 7. It means that the probability 
   of production of the pair of fake muons  with the
   charges of the opposite signs (like in signal 
   events) is  rather  high in these background events.
   Such pairs  may fake quite well the  signal events. 
   
   The distribution plots of the production vertexes of 
   background muons are shown in the right column of 
   Fig.{\bf 20}.  It is seen from the plot {\bf f} of  Fig.{\bf 20} that 
   the most of the muon  production vertexes are 
   spread over detector volume while for some of
   events these vertexes are rather close to the 
   interaction  point. So the information about the  vertex position 
   can be  useful for background separation.

\subsection{ Electron background}

~~~   Let us  consider now  the case of  electrons produced 
   in the background minimum-bias and QCD events. The
    distribution of the parents of  background electron 
   in the discussed above sample of minimum-bias and QCD 
   events is presented  in Fig.21 {\bf a}.  It is seen that, 
   like in a case of  fake electrons in signal events
   (see Fig. 17 {\bf a}), the main contribution comes
   from $\pi^{0}$-, $\eta$-, charged $K$-mesons   (bins
   2, 5 and 7, respectively) and also from muon decays
   (bin 1). The main source of fake electrons are   
   decays of neutral pions  ($\pi^{0} \rightarrow \gamma +  e^{+}e^{-}$). 
  They  can   appear directly or  from   decays of $\rho$-,
   $\eta$-,   $\omega$- , K- mesons,  as well as from 
   $\Delta$- resonances decays.

  From the Fig.21 {\bf b} one may  see that in the 
   minimum-bias and QCD events sample the structure of the 
   distributions of electron grandparents is rather
  \begin{figure}[!ht]
   \begin{center}
\vskip -0.5cm       
   \mbox{a)\includegraphics[width=7.2cm, height=5.2cm]{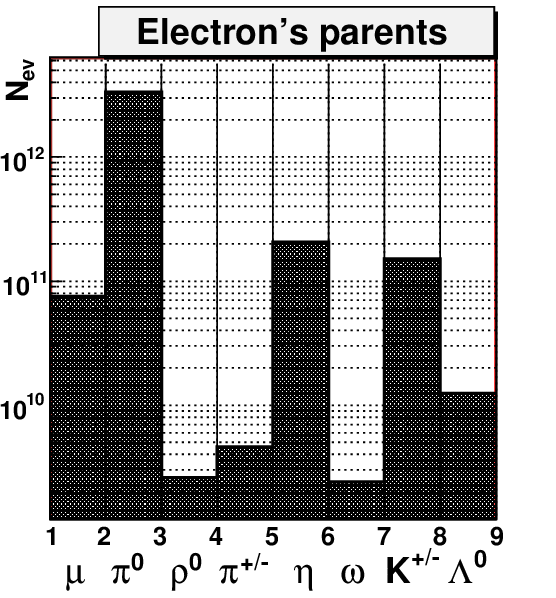}}  
   \mbox{b)\includegraphics[width=7.2cm, height=5.2cm]{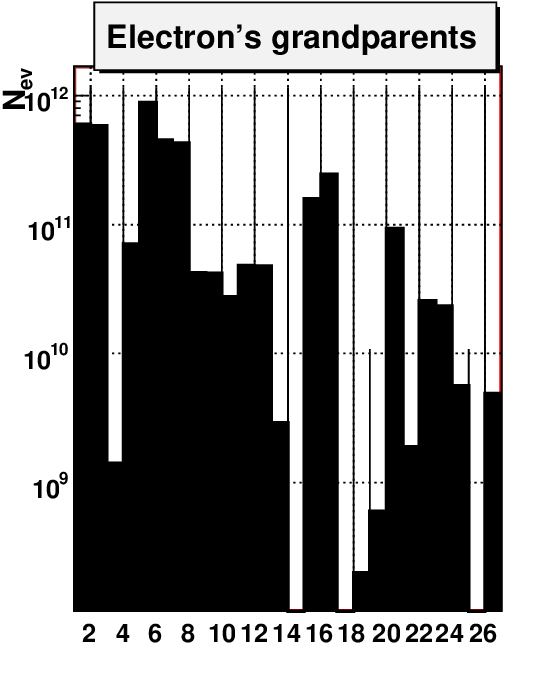}}  \\
     \caption{\small \it  Distributions of:
	       	 {\bf a)} parents of background $e^{+}/e^{-}$; 
	     {\bf b)} grandparents of background $e^{+}/e^{-}$ }		 
     \end{center}    
\vskip -0.5cm           
     \end{figure}  
   different from its analog presented  in Fig.17 {\bf  b} 
   for a case of fake electrons in the sample of signal
   events with the signal electron-positron pairs. Namely,
   different to the plot Fig.17 {\bf b}
   \footnote{after normalization to an equal number of 
              entries.}
   the charged
  $\rho$-mesons (bin 5) takes the dominant position 
   in the plot Fig.21 {\bf b}. Then follows the noticeably
   increased contribution of clusters (bin 1) which height
   reaches the height of strings contribution (bin 2). It 
   is also seen that on total the contribution of light vector 
   mesons ($\rho$, $\eta$ and $\omega$ (bins 5-7)), as  well
   as the contribution  of K- (bins 8-11) and $\eta^{'}$-, 
   $\phi$- mesons (bins 12,13), have grown up as comparing 
   to the higher bins (15, 16 and 20), which correspond
   to $\Delta$- and $\Sigma$- barions,  respectively.

  \begin{figure}[!ht]
     \begin{center}
\vskip -0.5cm       
   \mbox{a)\includegraphics[width=7.2cm, height=5.2cm]{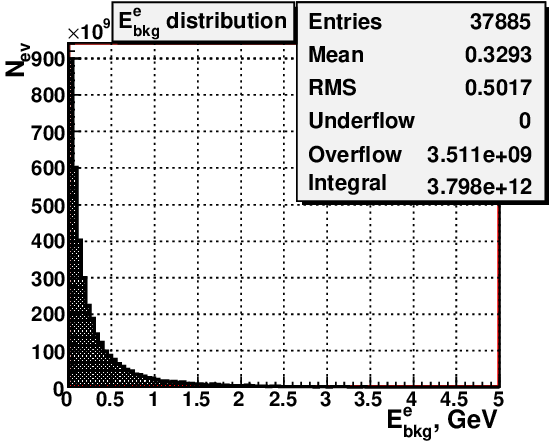}}  
   \mbox{b)\includegraphics[width=7.2cm, height=5.2cm]{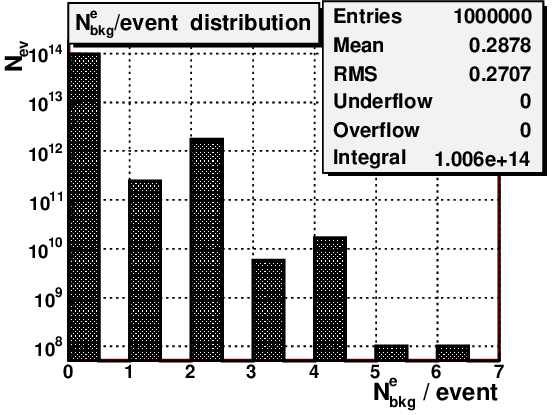}}  \\
   \mbox{c)\includegraphics[width=7.2cm, height=5.2cm]{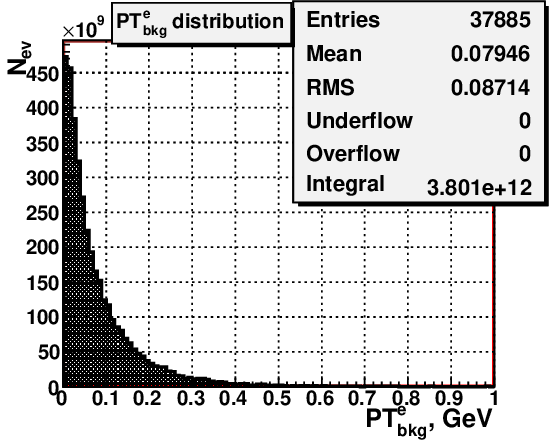}}  
   \mbox{d)\includegraphics[width=7.2cm, height=5.2cm]{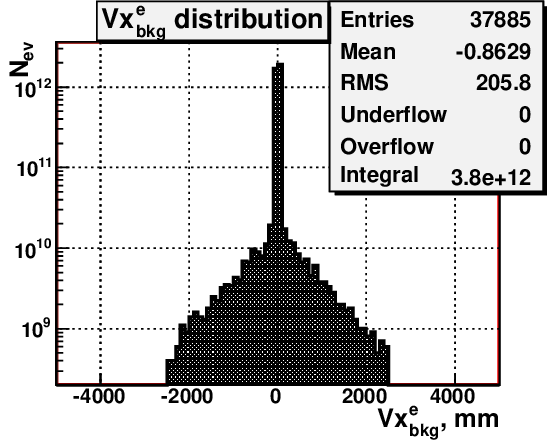}}  \\
   \mbox{e)\includegraphics[width=7.2cm, height=5.2cm]{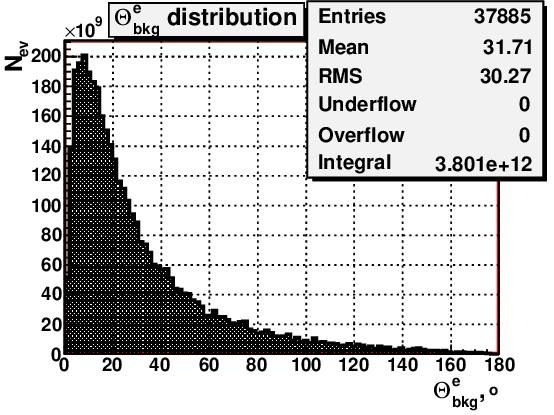}}  
   \mbox{f)\includegraphics[width=7.2cm, height=5.2cm]{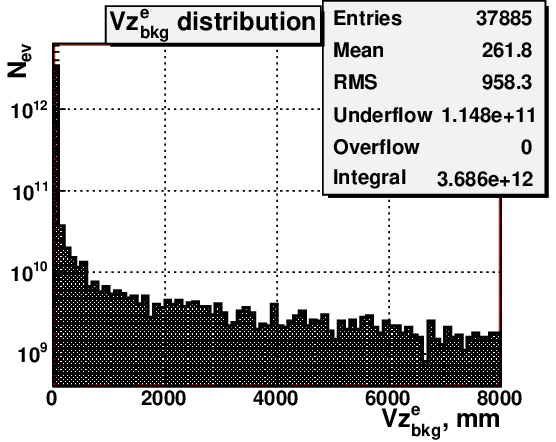}}  \\
     \caption{\small \it Left column:   
              distributions of the number of
              background electrons versus   {\bf a)}:
	       the energy $E^{e}_{bkg}$; 
             {\bf c)}:  transverse momentum  $PT^{e}_{bkg}$  and 
             {\bf e)}:   polar angle $\theta^{e}_{bkg}$.            
	       Right column: 
    distributions  {\bf b)}: of the total amount $N^{e}_{bkg}$  of electrons
              per the background event; 
	    {\bf d)}: x coordinate of the background $e^{+}/e^{-}$ production vertex
      and {\bf f)}: z coordinate of the background $e^{+}/e^{-}$ production vertex }
	 
     \end{center}    
\vskip -0.5cm           
     \end{figure}
    
   The distributions obtained from the sample of
   mentioned above generated  minimum-bias and
   QCD  background events, are shown in Fig.22.
   The  kinematical distributions (plots {\bf a, c} and {\bf e})
   are rather similar to the  distributions of fake 
   decay electrons in the signal events  which 
   were discussed in the subsection 6.4 and
   presented at Fig.18.
   From  Fig.22 one may see that the total number of    electrons
   in the  sample of generated 1000000 minimum-bias and
   QCD events   is equal to 37885. This number is
   of order less than the number of fake muons
   produced in the same sample of  $10^6$ minimum-bias 
   and QCD events (see previous
   subsection 7.1). 
   Plot {\bf b)} of  Fig.22 shows the distribution of the
   number of generated  minimum-bias and QCD
   events versus the number of  decay electrons per event. 
    The third bin in this plot,    as well as the other  bins
   to the  right from it, show how many events may contain
   two  and  more electrons. In these events there may 
   appear the $e^{+}e^{-}$-pairs, which potentially  may 
   fake  the signal events.  It is clearly seen that the probability
   of appearance of 2 and more electrons in the 
   final state  reduces to a value of about few percents 
   of the total  number of generated  events.

   The plots {\bf d} and {\bf f} of Fig.22 show the
   distributions of the position of the electron
   production vertex   in the background sample in the 
   transversal ($Vx^{e}_{dec}$) and the longitudinal
   ($Vz^{e}_{dec}$) directions. 
   It is  seen that the most of  background   electrons, originating 
   from  hadron decays, are produced near the interaction
    point ($V_{x}=0$ and $V_{z}=0$),  like it  take place  in Fig.18.

\section{ Background separation  }

~~~   To reduce the background contribution from
   the  minimum bias and  QCD events  we added 
    two new cuts to the previously used cuts 1)-3)
   (see Section 6.5). Finaly, we use the following  selection cuts:\\

$\bullet$ 
    1.) the events with the only  two leptons 
         with  $E_{l} \geq$ 0.2 GeV,  $PT_{l} \geq$ 0.2 GeV; 
	
$\bullet$
    2.)  the charges of these two leptons must   be of the opposite sign; 

$\bullet$  
  3.) the vertex of lepton origin lies within the radius $R_{vtx} \leq$ 15mm

   ~~~~~~  from the     interaction  point; 

$\bullet$   
    4.) $M_{inv} (l^+,l^-) \geq $ 0.9 GeV;

 $\bullet$ 
   5.) lepton isolation criteria: the  summed energy E$_{sum}$
         of all the particles around 
	
 ~~~~~the lepton within  the cone of  the radius 
      R=$\sqrt{\Delta^2_{\varphi} + \Delta^2_{\eta }}$ = 0.2
      in the $\eta -\varphi$ space
        
~~~~~is not higher  than E$^{max}_{sum} = 0.5$ GeV      
      \footnote{the  azimuth  angle $\varphi$ and the 
      polar (zenith) angle  $\theta$ 
      are used  to  determine the direction of the 
      3-momentum  of any  particle.  $\eta$ is  the 
      particle  pseudorapidity, defined  by the  
       formula  $\eta = -\ln  tg(\theta /2)$,
      where $\theta$ is  the  polar  angle
       of the particle 3-momentum counted from  the beam direction.}.\\
    
    Here $\Delta_\varphi= \varphi_{l} - \varphi_{p}$ is
    the difference of the lepton's (l) azimuth   angle  $\varphi_{l}$ 
    and the azimuth angle  $\varphi_{p}$ 
    of the particle (p), contained  within the
    cone  of the radius R around the lepton. Analogously, 
    $\Delta_{\eta }= \eta_{l} - \eta_{p}$  is the 
    difference of the lepton  and the particle
    pseudorapidities.

     Few words are in order now about the choice of
    the last two cuts. Tables 4 and 5 show,
     for muon pair and electron pair
    production cases, respectively, the influence of the 
    variation of the cut  on dilepton
    invariant mass  $M_{inv} (l^+,l^-)$ on the loss
    of signal events and the value of signal to
    background ratio S/B (after application of the
    cuts 1.)-3.). It is seen from Table 4 that
    in $\mu^{+}\mu^{-}$  case the growth of the 
    maximal value of $M_{inv} (l^+,l^-)$ up to
    $M_{inv} (l^+,l^-)= 1.2$ GeV allows to get rid 
    completely of the background  at  cost of
    loosing of 35 $\%$ of signal events, while in  
    $e^{+}e^{-}$ case, see Table 5, the same upper
    limit leads only to S/B$=$ 2.3.
        
\begin{table}[!hbt]
\vskip -0.5cm  
\caption{  Efficiency of the  $ M_{inv}^{ l^{+}l^{-} }$   cut for the 
case of $\mu^{+}\mu^{-} production$}
 \vskip -0.5cm    

\begin{center}
\begin{tabular}{|c|c|c|c|}              \hline 
$ M_{inv}^{ \mu^{+}\mu^{-} } \geq $ (GeV) & S/B for & Efficiency  & The rest of signal
 events  \\\hline
0.9  &    0.12  &  0.0805 &  85$\%$ \\\hline  
1.0  &    0.22  &  0.0444 & 82$\%$ \\\hline 
1.03 &    0.50  &  0.0198 & 81$\%$ \\\hline  
1.05 &    0.69  &  0.0136 & 78$\%$ \\\hline 
1.1  &    1.50  &  0.0063 & 76$\%$ \\\hline  
1.2  &   BKG = 0  &  0 & 65$\%$ \\\hline 
    
\end{tabular}
\end{center}
\vskip -0.5cm 
\end{table}

\begin{table}[!hbt]
\caption{  Efficiency of the  $ M_{inv}^{ l^{+}l^{-} }$   cut for the 
case of $e^{+}e^{-} production$}
 \vskip -0.5cm 
\begin{center}
\begin{tabular}{|c|c|c|c|}              \hline 
$ M_{inv}^{ e^{+}e^{-} } \geq $ (GeV) & S/B for & Efficiency  & The rest of signal
events  \\\hline
 0.9  &  0.093 &  0.0059  &   84$\%$ \\\hline  
 1.0  &  0.146 &  0.0035  &   81$\%$ \\\hline 
1.03 &  0.375 &  0.0014  &   78$\%$ \\\hline  
 1.05 &  0.395 &  0.0012  &   74$\%$ \\\hline 
1.1  &  0.762 &  0.0006  &   69$\%$ \\\hline  
 1.2  &  2.3   &  0.0002  &   61$\%$ \\\hline 
   
\end{tabular}
\end{center}
\vskip -0.5cm 
\end{table}

    The results of  all the five cuts sequent application
    to  the sample of inelastic $\bar{p}p \to X$ events  
    which contains the minimum-bias and QCD events (incuding
    the signal events based on the parton level annihilation
    subprocess ${q\bar{q} \to \gamma^{*} \to l^{+}l^{-}}$ )  
    are collected  in the  Table 6. It is seen that the first
    three cuts allow to enlarge the the S/B ratio by about
    one order in a case of muon pair production. At the
    same time, the third cut is inefficient for the case of 
    $e^{+}e^{-}$-pair production. The forth cut 
    $M_{inv} (l^+,l^-) \geq $ 0.9 GeV allows to increase
    the S/B ratio in $e^{+}e^{-}$ case  by more than two
    orders and by about three ordes the S/B ratio for 
    $\mu^{+}\mu^{-}$  case.

  \begin{table}[!hbt]  
 \vskip -0.5cm  
\caption{ Cuts influence for background events.}
 \vskip -0.5cm  
\begin{center}
\begin{tabular}{|c|c|c|c|c|}              \hline 
 N of cuts & S/B for $\mu^{+}\mu^{-} production$ & Efficiency  & 
 S/B for $e^{+}e^{-} production$  & Efficiency \\\hline
    1  &   1.41 $ \cdot 10^{-5}$  & 0.007  &  5.34 $\cdot 10^{-4}$ & 1.78 $\cdot 10^{-4}$  \\\hline  
    2  &   2.12 $\cdot 10^{-5}$  & 0.665  &  5.41 $\cdot 10^{-4}$ &  0.98  \\\hline  
    3  &   9.94 $\cdot 10^{-5}$  & 0.002  &  5.47 $\cdot  10^{-4}$ &  0.99     \\\hline  
    4  &   0.123                & 0.08   &  9.27 $\cdot  10^{-2}$ &  0.006     \\\hline  
    5  &   Background = 0       & --      &  3.8                 &  0.024     \\\hline 
     
\end{tabular}
\vskip -0.5cm  
\end{center}
\end{table}
 
   The plots presented in  the  Figs.23, 24 are done 
  to illustrate the action of the lepton isolation 
  criterion used in the definition of the fifth
  cut. They show the distributions of the total
  energy of the particles which are contained
  within the cones of the radius R around the 
  leptons. Figures 23 and 24 present these
  distributions for the  electron  and muon 
  cases, correspondingly. By comparing the plots
  {\bf a}  (for leptons from the signal events) 
  with the  plots {\bf b} (for background leptons
  from minimum-bias  and  QCD events) one can easily
  see  that the signal events have much smaller summarized energy content
  within the cone of $R \leq 0.2$ than the
  energy content in background events. This 
  observation is used in the cut 5.

\begin{figure}[!ht]
     \begin{center}
   \mbox{a)\includegraphics[width=7.2cm, height=5.2cm]{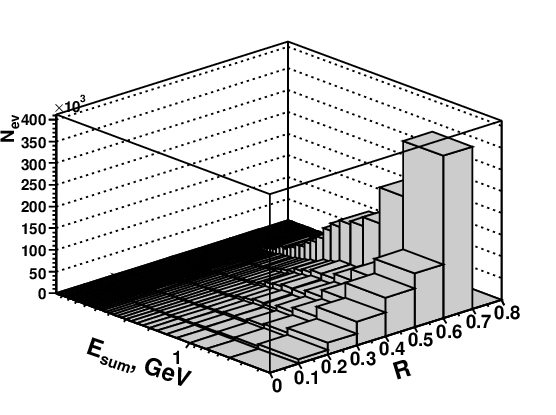}}  
   \mbox{b)\includegraphics[width=7.2cm, height=5.2cm]{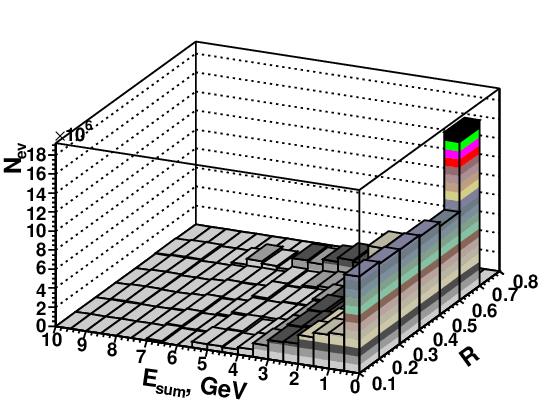}}  \\
     \caption{\small \it Lepton isolation for the e case.
     Distributions of the summarized energy $E_{sum}$ 
     within the cone of the radius R=0.1, 0.2, 0.3.... :
	       	 {\bf a)} signal events, 
	     {\bf b)} background events.}		 
     \end{center}    
\vskip -0.5cm           
     \end{figure}      
  
 \begin{figure}[!ht]
     \begin{center}
   \mbox{a)\includegraphics[width=7.2cm, height=5.2cm]{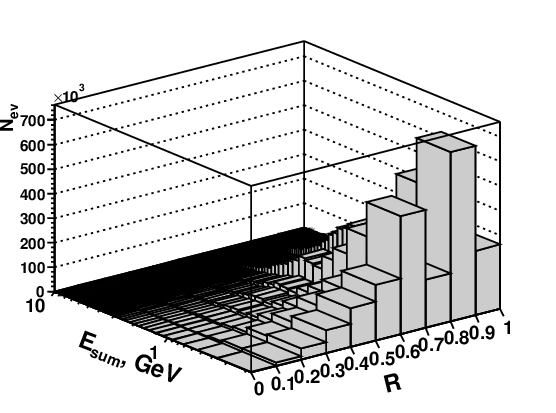}}  
   \mbox{b)\includegraphics[width=7.2cm, height=5.2cm]{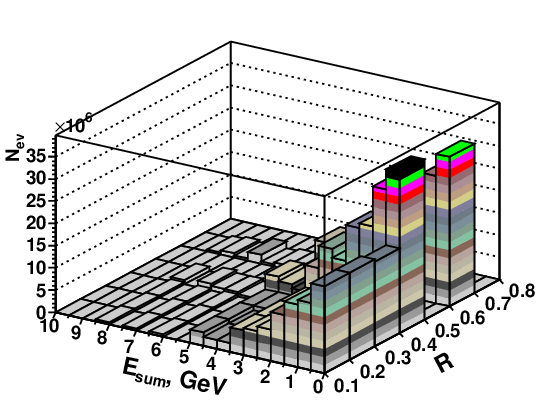}}  \\
     \caption{\small \it Isolation for the $\mu$ case.
      Distributions of the summarized energy $E_{sum}$ 
     within the cone of the radius R=0.1, 0.2, 0.3.... :
	       	 {\bf a)} signal events, 
	     {\bf b)} background events.} 		 
     \end{center}    
\vskip -0.5cm           
     \end{figure}

  From the Table 6 one can see that the last cut on 
  the lepton isolation, i.e. the choice of only those 
  final state leptons which have the restricted value 
  of the summarized energy (not greater than 
  $E_{sum}$=0.5 GeV) of other particles contained 
  within the  cone of some fixed radius
  R$=\sqrt{\varphi^2 + \eta ^2} = $NR
  (NR$ =0.1, 0.2, 0.3...$) around the direction
  of the lepton 3-momentum, allows one to achieve 
  (choosing NR$=0.2$) the value of the signal 
 to background ratio equal to S/B = 3.8 for  electron production case and 
completely to  get rid of background in muon production case. In both cases
   the application of the fifth cut
  leads to additional  8$\%$ loss of the signal
  events left after application of the first four cuts.  
  Let us note that the  same  criteria, but
  with the  use of a more restricted form of the
  forth cut $M_{inv} (e^+,e^-) \geq $ 1.0 GeV, 
  allows to increase the signal to background 
  ratio up  to S/B = 9  in  $e^+e^-$ case.


\section{ Remarks on the possibility of measuring the 
          multi-parton interactons and the intrinsic 
          quark transverse momentum in the proton.}
                               

~~~   In addition to the opportunity  to get the
  information about parton  distribution functions,
  which was already  discussed in Section 5,  let us also mention 
  here  three other ones which may be useful for studying
  quark  dynamics in proton and its PDFs.
   
  The first  two are  connected with the  processes  of two 
  \begin{equation}
   \bar{p}p \rightarrow l_{1}^{+}l_{1}^{-} + l_{2}^{+}l_{2}^{-} + X
   \end{equation}
  or even three lepton pairs 
   \begin{equation}
    \bar{p}p \rightarrow  l_{1}^{+}l_{1}^{-} + l_{2}^{+}l_{2}^{-} +
      l_{3}^{+}l_{3}^{-} + X
    \end{equation}
   production in one event.
   
   Both of these processes can include two  and three quark annihilation
   $q\bar{q} \rightarrow l^{+}l^{-}$  subprocesses, respectively.
   The total cross sections of such processes can be smaller as compared
   to the $ \bar{p}p \rightarrow l^{+}l^{-}  + X$
process which includes a single $q\bar{q} \rightarrow l^{+}l^{-}$ subprocess.
   Nevertheless, they may contain 
   interesting physical information which can
   be more easily extracted at the intermediate
   energies  than at the higher ones.  
   
    First, the measurement of the characteristics of the
   system of other than lepton pairs particles,
   produced in the process (4),
   will give us  the opportunity to get    the information about 
   the so-called "underlying" event. The
   study of the analogous distribution in the   process (5),  in 
   which all valence quarks (and
   antiquarks)  in proton (and antiproton) will 
   annihilate into lepton-antilepton pairs,
   may provide  the information about gluon content
   in the proton. The understanding of the physics of
   the "underlying" event, i.e. the interaction of
   partons which do not participate in the hard 
   subprocess $q\bar{q} \rightarrow l^{+}l^{-}$, is
   very important for the interpretation of the 
   results of the present Tevatron and future LHC
   experiments.
   
     The second  opportunity,   also provided by the processes 
   (4) and (5), is the study of the so-called multiple
   parton hard interaction processes in pp- and 
   $p\bar{p}$ interactions which are widely 
   discussed in connection with the problem of a 
   proper account of background contribution to the 
   processes which are planned for  seaches of
   the New Physics  signals at Tevatron and LHC. 
     It is worth mentioning that the
   measurements of the proceses (4) and (5) can
   be done for the case when final state lepton 
   pairs would have different flavor, like 
   $\bar{p}p \rightarrow e^{+}e^{-} + \mu^{+}\mu^{-} + X$.
   In such case we shall have the situation when 
   four hadronic jets events, used in previous
   measurements of multiple interactions \cite{AFS}-\cite{CDF93},
   are substituted by events withfour leptons. This substitution
   shall increase the  precision of measurement of
   the parameters of  multiple interections as it was 
   shown in recent measurements done with 
   "3jet + photon" final states \cite{CDF3jgam},
   \cite{D03jgam}.

    Besides getting the information about the
   fraction of multiple interactions their study
   opens the possibility to get the information about
   the spatial distribution of quarks within the
   proton. It is obvious that in a case of uniform 
   distribution of quarks within the proton volume the 
   occurrence of the first parton-parton interaction 
   would not influence on the probability of happening
   of the second interaction, while in a case when 
   the quarks are concentrated  in  small region
   the probability of happening of the second  
   interaction becomes higher if one of the quarks has 
   taken part in the first interaction.

     The third  opportunity is connected
  with the possibility to measure  the 
  characteristics of internal quark motion 
  in the proton. This possibility is based 
  on the fact that  the  shape of the distribution 
  of the modulus of the  vector sum of quark 
  and antiquark  transverse momentum vectors
  \begin{eqnarray}
    P^{q\bar{q}}_{T} = |\vec{P}^{q\bar{q}}_{T}| =
    |\vec{P}^q_{T} + \vec{P}^{\bar{q}}_{T} | 
  \end{eqnarray}
   practically  coincides 
  (due to the transverse momentum conservation 
  law) with the shape of the above-considered 
  distribution of the modulus of the  lepton pair 
  transverse  momentum   $P^{l^{+}l^{-}}_{T}$,
  shown in the plot {\bf {b}} of  Fig.12.
    \footnote{ Recall (see Section 4) that all
               plots in the  Fig.12 are done 
               for the case when  both
               ``Fermi motion''  (or "$k_{T}$-effect")  
               and ISR  are switched  ``on''.}

     The variable  $P^{q\bar{q}}_{T}$ is of  special
     interest because it contains the information about
     two important physical features of quark dynamics 
     inside the hadron.
     Indeed, in our case when a beam antiproton
     is directed along the z axis and it scatters 
     over the  proton  fixed target,
     there may be  only two  sources of  
     transverse motion of quarks in the     initial state 
\footnote {The values of transverse momenta 
           of constituents in a target proton
	   (which is at rest) as  well as of
	   those inside a beam antiproton 
           (which moves along the z axis) are
           invariant under Lorentz  boost  along the z axis.}:
 
{\bf  A)} internal Fermi-motion of  quarks
          (with some transverse   velocity) inside a 
          proton, i.e. the  so-called "$k_{T}$-effect";

{\bf  B)} initial-state radiation (ISR) of gluons or 
          photons from quarks before 
          hard quark-antiquark annihilation;

     The importance of these  two effects
    was recently  discussed  
    \cite{Huston} in connection with the
    interpretation of prompt  photon production
    study in the  experiment E706 at Fermilab
    \cite{E706}  and also with the  study of 
    "${\gamma/Z + jet}$" events,  which are 
    sensitive to the shape of gluon distribution,
    at the LHC \cite{BKSLHC} and Tevatron \cite{BSTevat},
    \cite {D0gam}.

%
\section{Conclusion.}
%
    
~~~~   The modeling of dilepton production in
  antiproton scattering over proton target
  $\bar{p}p \rightarrow l^{+}l^{-} + X$ is done
  for the intermediate energy $E_{beam}=14$ GeV 
  on the basis of PYTHIA event generator and the  
  parton level subprocess of quark-antiquark annihilation
  $\bar{q}q \rightarrow l^{+}l^{-}$.

    The distributions of most  essential kinematical
   variables of individual leptons,  are are  presented in  Section 2.
   They show   that the energy and  angle spectra of the fast (most 
   energetic) leptons in a pair  are very different from 
   those of slow leptons: the mean value 
   $ < {E^{l}_{fast}} > = 3.85$ GeV is about three 
   times higher than that  one of slow leptons
   $ < {E^{l}_{slow}} > = 1.36$ GeV. The simulation
   has  also shown a tendency which may be  a rather 
   general one:  fast leptons  fly predominantly at
   smaller angles $<{\theta^{l}_{fast}}> = 16.5^{o}$
   as  compared  to the angles of  slow 
   ones $<{\theta^{l}_{slow}}>$= $38.2^{o}$. It is 
   worth noting  that about  $6\%$ of events may 
   have slow leptons,   that may scatter into the back
   hemisphere, i.e. ${\theta^{l}_{slow}} > 90^{o}$.
   The angle-energy, energy-energy and angle-angle correlations
   among a slow and a fast lepton in the same lepton
   pair in  event are also described in Section 2 
   and are presented in  Figs.4, 5 together with the 
   corresponding distributions of the number of 
   events versus the corresponding lepton energies
   and angles. 
     
     These distributions  allow one
   to estimate the energy, transverse momentum and
   angle ranges that may be covered by leptons 
   produced in quark-antiquark annihilation process.
   They  were useful for proper design of muon system
   and may be also used for the electromagnetic
   calorimeter. Tables 1 and 2 show the estimation of
   the loss of signal events depending on the choise
   of cuts on the lower values of lepton energy and,
   respectively, on the upper limit of the angle
   size of the muon system and the electromagnetical
   calorimeter. From these Tables one can see that, for
   instance, the choise of cuts $E^{min}_{cut} = 0.5$ GeV
   and $\theta^{l}_{max} = 90^{o}$ results in 
   about 30$\%$ loss of signal events.   The simulation
    PYTHIA has shown that one may expect to gain
   about $7 \cdot 10^{7}$  MMTDY events per year for the
   luminosity $L = 2 \cdot 10^{5} mb^{-1}s^{-1}$.

   The analogous study was done on the basis of  PYTHIA in the Section 3
   for the leptons which may appear in decay of J/$\Psi$
   mesons, produced in the benchmark process 
   $\bar{p}p \rightarrow  J/\Psi + X$. It was shown  that
   the leptons, produced in J/$\Psi \rightarrow l^{+}l^{-}$ 
   decay, fit well into the same angle regions as
   the  leptons produced in MMTDY process 
   $\bar{p}p \rightarrow l^{+}l^{-} + X$. 
   The reconstruction of the lepton pair invariant mass can allow to
   get rid of background without a sizable loss of signal
   events. 

    In Section 4 the study of kinematical characteristics
   of lepton pair  as a whole system was done. It is  shown 
   that the spectrum of the invariant mass of the lepton 
   pair  decreases  rather fast and  vanishes at 
   $M^{l{+}l{-}}_{inv} = 2.5$ GeV.  At the same time  
   the lepton pairs  total energy  $E^{l^{+}l^{-}}$ 
   spectrum starts at around 1 GeV and extends  up to the 
   value of 12 GeV. It is also demostraited that about
   a  half of the events  have the  lepton pair energy 
   higher than 5 GeV.   It is shown that one can
    expect that in about 50$\%$ of  events the
   lepton pairs  would be  rather energetic   and they can take away
   from  33$\%$ up to 80$\%$ of the total energy of the
   final state system. The square of the 
   invariant mass  $(M_{inv}^{l^{+}l^{-}})^{2}=Q^2=q^2=
   ( P^{l^{+}} + P^{l^{-}})^2 $ has the meaning of the
   square of the momentum transfered from the hadronic
   system of quark-antiquark pair to the electromagnetic
   system of final state dilepton pair.

   In Section 5 the analysis of distributions, obtained
    by  PYTHIA, allowed to determine  the region in
   x-Q$^{2}$-plane which can be available for measuring 
    the  proton  structure function at PANDA. 
   This region  is  defined by the following boundaries:
   $0.05 \leq x \leq 0.7$ and  $Q^{2} \leq 6.25$ GeV.
   Let us emphasize that the measurements in this region
   of positive ("time-like")
   $q^2 \equiv ( P^{l^{+}} + P^{l^{-}})^2  = Q^2 \geq 0$
   would be a good extension of studies planned to be 
   done at JLab in the region of  negative ("space-like")
   values of  $q_{dis}^2=(P^{l}_{in} - P^{l}_{out})^2 \leq 0$.
        
    An important problem  of  background suppresion
  is considered in Sections 6, 7 and 8. In Section 6 
  we have concentrated  on the fake  leptons which  can  
  appear from hadron decays in the same signal process. 
  We studied signal processes with dimuon production
  separately from the processes with electron pair
  production using for this two separate
   event samples with signal muon pair
  and, respectively,  electron pair production. 
  First we have considered the case of
  $\mu^{+}\mu^{-}$ production. The histograms which 
  demonstrate the relative contribution of different
  parents and grandparents  of produced muons
  are presented in the subsection 6.1. It is shown 
  that charged pion decays  produce  the main 
  part of fake muons. The next dominant source is the
  decays of charged K-mesons (its contribution is
  by more than two ordes less than that one from
  pions). 

  Some details about pion source are presented
  in the subsection 6.2 where it was shown that about
  42$\%$ of signal events do not contain at all any
  charged pions, while 24$\%$ of them contain only
  one pion and 27$\%$ include two charged pion in 
  the final state. About 5$\%$ of signal events include
  tree charged pions and 1.5$\%$ have four charged
  pions. This prediction of PYTHIA indicate that the
  reconstraction of invariant masses of parent and
  grandparent hadrons can be a quite reliable way
  for fixing the origin of produced  fake  muons.

  The  kinematical plots, shown in subsection 6.3 
  (and in subsection 6.4 for fake electrons) are   done
  by applying a   geometrical restriction on the detector volume. 
  This restriction   produced  a strong reduction of the
  fraction of fake leptons.
  In result the energy and transverse 
  momentum spectra  of fake muons  became   essentialy shorter than their
  analogs  shown in the Section 2.
  The fraction of signal events which include fake
  muons has reduced down to about 16.6$\%$ (in comparison 
  with pion distributions). It means 
  that, according to PYTHIA,  about 83$\%$ of signal
  events  would not contain fake muons at all
   due to the used restriction of the decay volume. 
  The analogous parent, grandparent and kinematical
  distributions were obtained in subsection 6.4 for 
  a case of  background electrons.  It is found
  that  the application of  geometrical restriction on the detector volume 
  provides the  reduction  of signal events fraction
 (containing fake electrons) down  to 2$\%$.

   The set of three cuts are proposed in the   subsection 6.5
   which, being applied together with
   the mentioned above restriction on the detector
   volume, allows a further reduction of the fraction
   of the signal events containig fake decay leptons.
   Namely, they  are: $fr_{\mu}=0001\%$ in
   a case of  $\mu^{+}\mu^{-}$ production and
   $fr_{e}=0.008\%$ in $e^{+}e^{-}$. Let us underline
   that this strong reduction of the fraction 
   the signal events including fake leptons was
   achieved at the cost of the loss of a noticeable 
   number of selected signal events. These losses are: 
   $\approx 17\%$ for $\mu^{+}\mu^{-}$ and
   $\approx 14\%$ for $e^{+}e^{-}$  production.
   
   Much more dangerous background, provided  by  minimum-bias 
  and QCD events, was  studied in the Sections 7 and 8. The former
  includes two subsections in which the figures 
  with parent and grandparent relative contributions
  as well as the with kinematical distributions for
  $\mu^{+}\mu^{-}$ and $e^{+}e^{-}$ cases are 
  presented, respectively. Subsection 8 contains the
  set of five cuts which include three cuts which 
  were previously used to reduce the number of 
  events containing fake decay leptons. The fourth 
  cut reduces the spectrum of the invariant mass 
  of dilepton system by the condition
  $M_{inv}(l^{+},l^{-}) \geq 0.9$ GeV, while the
  fifth  cut uses  the isolation criteria for the
  lepton. It is shown that dispite the fact that
  the cross section of minimum-bias process 
  $\sigma^{minbs}_{tot}$ is by about 7 oders 
  larger than the cross section of the signal process 
  $\sigma^{\bar q q \to l^{+}l^{-}}_{tot}$
  the  application of these five cuts allows 
  to get rid completely of minimum-bias and QCD background
  contribution in $\mu^{+}\mu^{-}$case  and to 
  reach the value of S/B =  3.8 for $e^{+}e^{-}$  case.
   It is worth mentioning that the application of the
  fourth and the fifth cuts leades to an additional loss of
  the number of selected signal events by 8$\%$.

   The Section 9 contains three important 
  remarks about the physical potential of studing
  events having several lepton-antilepton pairs
  in the final state.
  First, it is stressed that the  study of events 
  with two (and, maybe, even three) lepton pairs
  would allow to enlarge the precision of the
  parameters of multiple quark interactions. It 
  will also essentialy extend the
  region of QCD studies because up to now there
  were done only five dedicated measurements of
  such events in proton-proton and 
  antiproton-proton collisions. The first three
  processes considered the case when four jets
  were produced in the final  state, while
  the last two measurements have  used the events in
  which the final state  was including
   tree jets plus one direct photon.   Therefore, the
  measurements in which the jets would be 
  substituted by leptons would allow to reach
  a  higher level of precision.

     It was also noted that at the same 
  time such events, based on the
  processes of valence quarks annihilation in
  the colliding protons, will provide a clean
  information about the dynamics of spectator 
  quarks interactions, i.e. about the so-called
  "underlaying events". It is worth mentioning
  that the  most interesting would be the 
  measurements with three leptons pairs 
  production, because in this case the 
  underlaying processes  would be defined
  mostly by soft gluon interactions.

    The third opportunity, which was discussed 
   in the Section 8,  can be based on the
   measurement of the transverse momentum 
   of lepton pair which is directly connected 
   with the transverse momentum  of the system
   of two annihilating quarks. The latter can be
   caused by the so-called "$k_{T}-effect$" which
   is connected with the so called "Fermi-motion"
   of quarks inside the
   proton or the radiation of gluons by initial-state
   quarks. This information is of a big interest for
   the interpretation of QCD effects observed 
   at high energy hadron colliders. 

    It should be underlined  that the present PYTHIA
   simulation does not take  into account the detector
   effects (like the  magnetic field and the material
   of the apparatus, for example). 
   Hence, the obtained plots, 
   which describe  the unbiased distributions of the 
   produced free particles, may be  mainly useful for 
   preliminary estimations and working out the criteria
   (cuts) for selection of experimental events for 
   further physical analysis. 
   The detailed GEANT simulation
   with account of  detector design (and based on the 
   simulated PYTHIA event sample  described here)
   will  be a subject of our following publication.
   Nevertheless, one can expect that there is a high 
   probability that the main  features of  the real
   process, corrected to detector effects, would be 
   rather similar to those shown in the plots
   presened in this article.

\section{Acknowledgements.}

~~~One of the authors (A.N.S.) acknowledge support from Russian-German ''FAIR Russia Research Centre`` (which is also supported from Russian Federal Agency for Atomic Energy ''Rosatom`` and German Helmholtz Association).




\newpage 
   \section{Appendix: Tables.} 

\begin{table}[hb]
\caption{Bin correspondence to the particle names for the
    case of {\it \bf e}'s and {\it \bf ${\mu}$}'s
    grandparents.}
\begin{center}
\begin{tabular}{|c|c|c|}            \hline 
 Bin number & Name of {\it \bf e}'s grandparent particle   
            & Name of {\it \bf$\mu$}'s grandparent partticle  \\\hline
    1       &    cluster        &  cluster        \\\hline  
    2       &    string         &  string         \\\hline    
    3       &    $\rho^0$       &  $\rho^0$       \\\hline  
    4       &    $\pi^+$        &  $\rho^+$       \\\hline  
    5       &    $\rho^+$       &  $\eta$         \\\hline  
    6       &    $\eta$         &  $\omega$       \\\hline  
    7       &    $\omega$       &  $K^0_S$        \\\hline  
    8       &    $K^0_S$        &  $K^{*0}$       \\\hline  
    9       &    $K^{*0}$       &  $K^+$          \\\hline   
    10      &    $K^+$          &  $K^{*+}$       \\\hline  
    11      &    $K^{*+}$       &  $\eta'$        \\\hline       
    12      &    $\eta'$        &  $\phi$         \\\hline   
    13      &    $\phi$         &  $\Delta^-$     \\\hline 
    14      &    $\Delta^-$     &  $\Delta^0$     \\\hline  
    15      &    $\Delta^0$     &  $K^{+/-}$ like string   \\\hline   
    16      &    $\Delta^+$     &  $\Delta^+$     \\\hline  
    17      &    $K^{+/-}$ like string   &  $\Delta^{++}$  \\\hline  
    18      &    $\Delta^{++}$  &  $\Sigma^{-}$   \\\hline  
    19      &    $\Sigma^{*-}$  &  $\Sigma^{*-}$  \\\hline  
    20      &    $\Lambda^0$    &  $\Lambda^0$    \\\hline  
    21      &    $\Sigma^{0}$   &  $\Sigma^{*0}$  \\\hline  
    22      &    $\Sigma^{*0}$  &  $\Sigma^{+}$   \\\hline  
    23      &    $\Sigma^{+}$   &  $\Sigma^{*+}$  \\\hline  
    24      &    $\Sigma^{*+}$  &  $\Xi^-$        \\\hline  
    25      &    $\Xi^{*-}$     &  $\Xi^{*-}$     \\\hline  
    26      &    $\Xi^{0}$      &  $\Xi^{*0}$     \\\hline   
\end{tabular}
\end{center}
\end{table}

\newpage 
\begin{table}[hb]
\caption{ Bin correspondence to the
 particle names for the case of  $\pi^{+/-}$ 's
  parents and grandparents. 
}
\begin{center}
\begin{tabular}{|c|c|c|}              \hline
 Bin number & Name of {\it \bf ${\pi}$}'s parent particle & 
              Name of {\it \bf ${\pi}$}'s grandparent particle \\\hline
    1       &    cluster        &  d - quark         \\\hline  
    2       &    string         &  u - quark         \\\hline  
    3       &    $\rho^0$       &  cluster           \\\hline  
    4       &    $\rho^+$       &  string            \\\hline  
    5       &    $\eta$         &  $\rho^0$          \\\hline  
    6       &    $\omega$       &  $\omega$          \\\hline  
    7       &    $K^0_S$        &  $K^0$             \\\hline  
    8       &    $K^{*0}$       &  $K^{*0}$          \\\hline  
    9       &    $K^+$          &  $K^{*+}$          \\\hline  
    10      &    $K^{*+}$       &  $\eta'$           \\\hline  
    11      &    $\eta'$        &  $\phi$            \\\hline  
    12      &    $\phi$         &  $ud-diquark, S=0$    \\\hline  
    13      &    $\Delta^-$     &  $ud-diquark, S=1$    \\\hline  
    14      &    $\Delta^0$     &  $uu-diquark, S=1$    \\\hline  
    15      &    $\Delta^+$     &  neutron           \\\hline  
    16      &    $\Delta^{++}$  &  $\Sigma^{*-}$     \\\hline  
    17      &    $\Sigma^{-}$   &  $\Sigma^{0}$      \\\hline  
    18      &    $\Sigma^{*-}$  &  $\Sigma^{*0}$     \\\hline  
    19      &    $\Lambda^0$    &  $\Sigma^{*+}$     \\\hline  
    20      &    $\Sigma^{*0}$  &  $\Xi^{-}$         \\\hline  
    21      &    $\Sigma^{+}$   &  $\Xi^{*-}$        \\\hline  
    22      &    $\Sigma^{*+}$  &  $\Xi^{0}$         \\\hline  
    23      &    $\Xi^-$        &  $\Xi^{*0}$        \\\hline  
    24      &    $\Xi^{*-}$     &                    \\\hline 
    25      &    $\Xi^{*0}$     &                    \\\hline  
	      
\end{tabular}
\end{center}
\end{table}



\begin{thebibliography}{50}


\bibitem{Matv}
V.A.~Matveev, R.M.~Muradian, A.N.~Tavkhelidze,
JINR P2-4543, JINR, Dubna, 1969;
SLAC-TRANS-0098, JINR R2-4543, Jun 1969; 27p. 

\bibitem{Drell}
S.D.~Drell, T.M.~Yan, SLAC-PUB-0755, Jun 1970, 12p.;  Phys. Rev. Lett.
{\bf 25}(1970)316-320, 1970

\bibitem{DYexper} 
CERN UA1 Collaboration, C.~Albajar et al., 
Phys. Lett., { \bf B209} (1988) 397;
FNAL E772 Collaboration, P.L.~Gaughey et al.
Phys. Rev. { \bf D50} (1994) 3038

\bibitem{MRST}
A.D.~Martin  et al., Eur. Phys. J. {\bf C4} (1998) 463

\bibitem{PANDALoI}
``Strong interaction studies with antiprotons. Letter of intent
for PANDA (antiproton annihilations at Darmstadt)''.
By PANDA Collaboration (M. Kotulla et al.). Jan 2004. 34pp.
Electronic Version from a server  EXP GSI-FAIR-PANDA 

\bibitem{PANDATPR}
``Strong Interaction Studies with Antiprotons. Technical
Progress Report for PANDA; FESR-ESAC/Pbar/Technical
Progress Report, by PANDA Collaboration, GSI, 2005

\bibitem{PANPysBook} "Physics Book of PANDA''
 by PANDA Collaboration, GSI, 2007
 
\bibitem{ASSIA}
M.Maggiora  et al., ``Spin physics with antiprotons'',
hep-ph/0504011, 2005; \\
V.Abazov  et al., ``A study of spin-dependent interactions 
with antiprotons: the structure of the nucleon.''
hep-ph/0507077, 2005

\bibitem{PAX}
F.Rathmann, P.Lenisa,  Spin  physics at GSI,
hep-ex/0412078, 2004; \\
V.Barone  et al., Antiproton-proton scattering 
experiments with polarization,
hep-ph/0505054, 2005




\bibitem{R.Baier} 
R.Baier and R.R\"{u}ckl, Z.Phys. {\bf C19} (1983) 25

\bibitem{M.Drees}
   M.Drees and C.S.Kim, Z.Phys. {\bf C53} (1991) 673

\bibitem{G.T.Badwin}
G.T.Bodwin, E.Braaten and G.P.Lepage, Phys. Rev. {\bf D51} (1995) 1125 
[Erratum: ibid  {\bf D55} (1997) 5883];
M.Beneke, M.Kr\"{a}mer and M.V\"{a}nttinen, Phys. Rev. {\bf D57} (1998) 4258;
B.A.Kniehl and J.Lee, Phys. Rev. {\bf D62} (2000) 114027


\bibitem{Note1}
A.N.Skachkova, N.B.Skachkov,
``Muon pair production in proton-antiproton interactions
at intermediate energies'',
hep-ph/0412279, 2004

\bibitem{Note2}
A.N.Skachkova, N.B.Skachkov,
``Monte-Carlo simulation of lepton pair production in
$\bar{p}p \rightarrow l^{+}l^{-} + X$ events at 
$E_{beam}=14$ GeV'', hep-ph/0506139, 2005

\bibitem{Sjost}
T.Sj\"{o}strand, S.Mrenna, P.Skands, JHEP 05(2006) 026;
hep-ph/0603175, LU TP 06-13; FERMILAB-PUB-052-CD-T, March 2006


\bibitem{And79}
B. Andersson, G. Gustafson and C. Peterson,
Z. Phys.  { \bf C1} (1979) 105;

B. Andersson, G. Gustafson, 
Z. Phys.  { \bf C3} (1980) 22;

B. Andersson, G. Gustafson and T. Sj\"{o}strand,
Z. Phys.  { \bf C6} (1980) 235; Z. Phys.  { \bf C12} (1982) 49

 
\bibitem{And80} 
B. Andersson, G. Gustafson  and T. Sj\"{o}strand,
Phys. Lett. { \bf B94} (1980) 211

\bibitem{And81}
B. Andersson, G. Gustafson, I. Holgersson and O. Mansson,
Nucl. Phys. { \bf B178} (1981) 242

\bibitem{And81a}
B. Andersson, G. Gustafson, G.Ingelman and T. Sj\"{o}strand,
Z. Phys.  { \bf C9} (1981) 233

\bibitem{And82}
B. Andersson, G. Gustafson and T. Sj\"{o}strand,
Nucl. Phys. { \bf B197} (1982) 45

\bibitem{And83}
B. Andersson, G. Gustafson, G. Ingelman and T. Sj\"{o}strand,\\
Phys. Rep.  { \bf 97} (1983) 31

\bibitem{AFS}
T. Akesson et al. (AFS Collab.), Z.Phys. C {\bf 34} (1987) 163

\bibitem{UA2}
J. Alitti et al. (UA2 Collab.), Phys. Lett. B {\bf 268} (1991) 145

\bibitem{CDF93}
F. Abe et al. (CDF Collab.) Phys. Rev. D {\bf 47} (1993) 4857

\bibitem{CDF3jgam}
F. Abe et al. (CDF Collab.) Phys. Rev. Lett. {\bf 79} (1997) 584,
Phys. Rev. {\bf D  56} (1997) 3811,

\bibitem{D03jgam}
V. M. Abazov et al. (D0 Collab.)  Phys. Rev. {\bf D  81} (2010) 052012

\bibitem{E706}
Fermilab E706 Collab., L. Apanasevich et al., Phys. Rev. {\bf D70}:092009,
2004, hep-ex/0407011;  Phys. Rev. Lett. 81:2642-2645, 1998,  
hep-ex/9711017

\bibitem{SBKLHC}
N.B.~Skachkov, V.F.~Konoplyanikov D.V.~Bandourin,
Third Annual RDMS CMS Collaboration Meeting.
CMS-Document, 1997--168. CERN, December 16-17, 1997,
p.139-153

\bibitem{BKSLHC}
D.V. Bandurin, V.F. Konoplyanikov, N.B. Skachkov, hep-ex/0207028

\bibitem{BSTevat}
D.V. Bandurin, N.B. Skachkov, D0-NOTE-3948, FNAL, Mar.2002, 
hep-ex/0203003, 2002;  Phys. Part. Nucl. 35:66-106, 2004
(Fiz. Elem. Chast. Atom. Yadra 35:113-177,2004), hep-ex/0304010 
  
 \bibitem{D0gam}
 V. M. Abazov et al. (D0 Collab.), Phys. Lett. {\b B  666} (2008) 435; 
  Phys. Rev. Lett.  {\bf D  102} (2009) 192002

\bibitem{ROOT}
R.~Brun and F. Rademakers, 
ROOT - An Object Oriented Data Analysis Framework, 
Proceedings AIHENP '96 Workshop, Lausanne, Sep. 1996,
Nucl. Inst. \& Meth. in Phys. Res. 
{ \bf A389} (1997) 81.
See also http://root.cern.ch/


\bibitem{Huston}
L. Apanasevich et al., Phys. Rev. { \bf D59}:074007,1999; hep-ph/9808467;\\
J. Huston; Int. J.Mod.Phys. A16S1A:205-208,2001; 

%
\end{thebibliography}
\end{document}